\documentclass[aps,prd,showkeys,showpacs,amssymb,eqsecnum,amsfonts,
preprintnumbers,nofootinbib,superscriptaddress]{revtex4}

\usepackage{graphicx}
\usepackage{natbib}
\usepackage{latexsym}
\usepackage{amsfonts}
\usepackage{bm}

\begin{document}

\newcommand{\vp}{\varphi}
\newcommand{\nn}{\nonumber\\}
\newcommand{\beq}{\begin{equation}}
\newcommand{\eeq}{\end{equation}}
\newcommand{\bed}{\begin{displaymath}}
\newcommand{\eed}{\end{displaymath}}
\def\bea{\begin{eqnarray}}
\def\eea{\end{eqnarray}}

\title{Casimir effect in a 6D warped flux compactification model}
\author{Masato~Minamitsuji}
\email[Email: ]{masato"at"theorie.physik.uni-muenchen.de}
\affiliation{Arnold-Sommerfeld-Center for Theoretical Physics, Department f\"{u}r Physik, Ludwig-Maximilians-Universit\"{a}t, Theresienstr. 37, D-80333, Munich, Germany}

\begin{abstract}
We discuss Casimir effect of a massless, minimally coupled scalar field
in a 6D warped flux compactification model and its implications for the 
hierarchy and cosmological constant problems, which
are longstanding puzzles in phenomenology and cosmology.
Due to the scale invariance of the background theory, the 4D effective theory
contains a volume modulus.
To stabilize the modulus, we consider one-loop corrections to the effective
potential by the Casimir effect.
The one-loop effective potential for the volume modulus 
has a form which is very similar to Coleman-Weinberg potential.
We evaluate coefficients appearing in the effective potential by
employing zeta function regularization and heat kernel analyses.
The volume modulus is stabilized for smaller degrees of warping, below a 
critical value, which depends on deficit angle of the reference brane. 
After stabilizing the modulus, it is possible to obtain observed
values of the mass ratio between the fundamental energy scales and a tiny effective cosmological constant (though its sign is negative).
The degree of warping should be tuned to be close to the critical value,
not as severely as the original fine-tuning.
\end{abstract}

\preprint{LMU-ASC 28/07}
\date{\today}
\maketitle

\section{Introduction}

There are several longstanding problems in phenomenology and cosmology.
One of them is why gravity is so weak in
comparison with the electroweak interactions
(in other words why Planck scale $M_{\rm Pl}\sim 10^{19}{\rm GeV}$ is much
larger than that of electroweak interaction $M_{\rm EW}\sim 10^{3}{\rm GeV}$)
and is known as the {\it hierarchy problem}.
Another important problem is why 
the energy density of dark energy
which dominates the present Universe
(assuming that its origin is vacuum energy of quantum fields) 
$\rho_{\rm Vac}\sim (10^{-3}{\rm eV})^4$ is 
much smaller than that of gravitational scale $M_{\rm Pl}^4$,
which is naturally expected from the standard model and
is known as the {\it cosmological constant problem}.

In this article, we focus on the Casimir effect in a 6D braneworld model whose
extra-dimensions are compactified by magnetic flux \cite{Gibbons:2003di}.
6D flux compactification models have attracted much attensions because
of the fascinating feature that they may help resolve the above
puzzles, namely the hierarchy problem \cite{Arkani-Hamed:1998rs} 
\footnote{For extensions of the model given in \cite{Arkani-Hamed:1998rs} 
with more successful localizations
of the standard model particles on intersections of D-branes,
see e.g., \cite{Kokorelis}.}
and the cosmological constant problem \cite{Carroll:2003db}.
It is also expected that in these models effects of extra-dimensions  
may be detected in future experiments on gravity
because they may become important at sub-milimeter scales
\cite{Arkani-Hamed:1998rs},
where Newton's law has not been strictly confirmed.

The basic motivation to consider the Casimir effect in such a 6D brane model
is as follows. An important point is that $M_{\rm Pl}$ and $M_{\rm EW}$
have satisfy $M_{\rm Pl}\sim 10^{16} M_{\rm EW}$ and
$\rho_{\rm Vac}^{1/4} \sim 10^{-16} M_{\rm EW}$.
Thus in these two hierarchies there is substantial similarity
and it could be naturally expected that if there is a theory which gives a common factor of $10^{16}$, both hierarchy problems may be solved at the same time. The Casimir effect in 6D spacetime may be able to give such 
a common factor.
\footnote{After the proposal of Randall-Sundrum braneworld model
\cite{Randall:1999ee}, the Casimir effect in 5D braneworld has also been discussed
in terms of phenomenology, especially toward a resolution of the
hierarchy problem, see e.g.,~\cite{Garriga:2000jb}.}
Let's imagine a spacetime with $n$ extra dimensions
(later we set $n=2$) whose size is assumed to be stabilized at 
a characteristic scale $a$.
Then, the dimensionaly reduced Planck mass is given by
$M_{\rm Pl}^2=a^n M_{4+n}^{2+n}$ \cite{Arkani-Hamed:1998rs},
where we assume $M_{4+n}\sim M_{\rm EW}$.
Casimir energy density induced on the brane by fields living
in the internal space is roughly given by $\rho_{\rm Cas}\sim a^{-4}$.
Thus, we get
\begin{eqnarray}
\frac{\rho_{\rm Cas}}{M_{4+n}^4}
\sim \left(\frac{M_{4+n}}{M_{\rm Pl}}\right)^{8/n}\,.
\label{ever}
\end{eqnarray}
Especially, for the case of $n=2$, we obtain
$\rho_{\rm Cas}/M_{6}^4 \sim (M_6/M_{\rm Pl})^{4}$.
If the Casimir energy density $\rho_{\rm Cas}$ plays the role of 
the dark energy density $\rho_{\rm Vac}$, 
we can get the desired ratio (see also \cite{Chen:2006nu}).
The problem is whether the ratio is really obtained from the 
setup of the 6D braneworld.
Bearing the above considerations in mind,
in this article, we focus on the Casimir effect in a specific model
of 6D braneworld with a warped flux compactification.
\footnote{In the above discussion here, we implicitly assume that 
the tree level vacuum energy density is effectively
cancelled by some self-tuing mechanism \cite{Carroll:2003db}, though 
there are some criticisms for this mechanism \cite{cc2}.
We just discuss the impact of the Casimir effect (in a 6D model) on the cosmological constant problem and implicitly assume that a self-tuning mechanism exists.}

We analyze the Casimir effect induced by a massless minimally coupled scalar field
in a warped flux compactification
model based on a 6D (Salam-Sezgin) supergravity
\cite{Nishino:1984gk,Salam:1984cj}, employing zeta function regularization
techniques. In the 4D effective theory of the model, a volume modulus arises due to the scale invariance of the 6D theory and is stabilized
by one-loop quantum corrections to the effective potential
induced by the bulk scalar field.
We then discuss implications for the hierarchy and cosmological constant problems. In Ref. \cite{mns, emn} the Casimir effect in a 4D toy model, which has 
a very similar spacetime structure to this 6D model, has been discussed. 
As a consequence, it has been suggested that for larger degrees of warping and smaller degree of dilaton coupling, the Casimir effect may give a resolution to  the hierarchy and cosmological constant problems.
So, it is desirable to discuss the Casimir effect in the original 6D model.
In attempting to evaluate the Casimir effect in the original 6D  model, however,
there is a technical problem arising from a lack of mathematical formulation of conical heat kernel coefficients (in 6D), which are relevant to  
contributions of the boundary 3-branes to the effective potential.

To overcome the above technical problem, 
we instead focus on a special relation 
between the relevant heat kernel coefficient and
the analytically continued zeta function \cite{VSV}.
The mass spectrum which includes brane contributions is now available
in a conformally related (unwarped) spacetime and
is very similar to the case of the 4D model.
Thus, instead of trying to evaluate the relevant heat kernel coefficient,
we shall use the analytically continued zeta function
by employing the mode sum technique,
which has been developed in the recent work on the 4D toy model, Ref. \cite{emn}.

The article is organized as follows. In Sec. II, we briefly introduce the background model for a 6D warped flux compactification.
In Sec, III, we discuss the quantum mechanical perturbations of a massless
minimally coupled scalar field. In Sec. IV, we derive mathematical relations 
in order to analyze the stability of the volume modulus and the Casimir effect.
In Sec. V, we will discuss volume stabilization and its implications for the Casimir effect on the hierarchy and cosmological constant problems.
In Sec VI, we shall close this article after giving a brief summary
and discussions on possible extensions of the present work.

\section{A warped codimension two brane model with flux compactification}
 
 \subsection{Solution}
 
We consider a 6D Einstein-Maxwell-dilaton theory with a non-vanishing scalar potential \cite{Nishino:1984gk,Salam:1984cj,Gibbons:2003di} as
\begin{eqnarray}
  S_6=
M_6^4
\int d^6 x\sqrt{-g}
\left(
 \frac{1}{2}R
-\frac{1}{2}\partial_A \varphi \partial^A \varphi
-\frac{1}{4} e^{-\varphi}F_{AB}F^{AB}
 -2g^2 e^{\varphi} 
\right)\,,
\label{theory6d}
\end{eqnarray}
where $\varphi$ is a dilaton field, $F_{AB}$ represents a $U(1)$ gauge field strength and $g$ is the $U(1)$ gauge coupling constant.
This theory corresponds to the bosonic part of the Salam-Sezgin,
6D supergravity \cite{Salam:1984cj, Nishino:1984gk}. 
Hereafter we set $M_6^4=1$ for simplicity and if needed, we
put it back in explicitly.  

This theory contains a series of solutions of warped flux compactification
 \cite{Gibbons:2003di} ;
\begin{eqnarray}
ds^2&= & h(\rho)d\theta^2
     +\frac{d\rho^2}{h(\rho)}
     +(2\rho)\eta_{\mu\nu}dx^{\mu}dx^{\nu}\,,
     \nonumber\\
     &&h(\rho)=\frac{g^2}{2\rho^3}
       \left(\rho_+^2 -\rho^2\right)
       \left(\rho^2 -\rho_-^2\right)\,,
     \nonumber\\
     &&\varphi(\rho)=-\ln(2\rho)\,,
     \nonumber\\
     &&F_{\theta\rho}= -\frac{g\rho_+\rho_-}{\rho^3}\,,
\label{metric}
\end{eqnarray}
where two 3-branes are located at $\rho=\rho_{\pm}$.
For later convenience, we define a new parameter
\begin{eqnarray}
\alpha=\frac{\rho_-}{\rho_+}\,,
\end{eqnarray}
which controls shape (i.e., warping) of the internal space.

To see the spacetime structure in details, it is useful to introduce
a new coordinate as
\begin{eqnarray}
z= \left(\frac{\rho_+^2-\rho^2}{\rho^2-\rho_-^2}\right)^{1/2}\,,
\quad
\varphi=\frac{g^2(1-\alpha^2)}{2\kappa} \theta\,,
\label{mitsuo}
\end{eqnarray}
where $\kappa=1-\frac{\delta_+}{2\pi}$.
\begin{eqnarray}
ds^2= (2\rho)
 \left[
 \frac{dz^2}{g^2 (1+z^2)^2}
+z^2 \frac{\kappa^2d\varphi^2}{g^2 (1+\alpha^2z^2)^2 }
+\eta_{\mu\nu}dx^{\mu}dx^{\nu}
\right]\,.
\label{bulkmetric}
\end{eqnarray}

The braneworld action is given by
\begin{eqnarray}
 S_{\pm}
=-\int d^4 x \sqrt{h}
 \sigma_{\pm}\,,
\end{eqnarray}
respectively, where $\sigma_{\pm}$ denotes the brane tensions,
which are related to to the conical deficit angles by
\begin{eqnarray}
\sigma_{\pm}=  M_6^4 \delta_{\pm}\,.
\end{eqnarray}
$h_{ab}$ is the brane induced metric.
In order to share the same angle period around both the branes,
the deficit angles are related to $\alpha$ as 
\begin{eqnarray}
 \frac{2\pi-\delta_+}{2\pi-\delta_-}
  =\alpha^2\,. \label{tendef6d}
\end{eqnarray}
Eq. (\ref{tendef6d}) implies that once the brane tensions, $\sigma_{\pm}$
are specified,
then the bulk shape $\alpha$ is also fixed.
We now regard $\alpha$ and $\kappa$ as free parameters,
instead of $\sigma_{\pm}$,
along with $g$.
So we shall use $(+)$-brane as a reference brane.
The remaining modulus is the absolute size of the bulk, $\rho_+$. 
Note, however, that there is also magnetic flux constraint given by
\begin{eqnarray}
 \int^{\rho_+}_{\rho_-} d\rho
 \int^{\Delta\theta}_0 d\theta
 F_{\rho\theta}
 = -\Delta\theta \times(g\rho_+\rho_-)
   \big(\frac{1}{\rho_-^2}-\frac{1}{\rho_+^2}\big)
 =-\frac{4\pi\kappa}{g \alpha}  \,,
\end{eqnarray}
and the magnetic flux only depends on $\alpha$ and $\kappa$ and does not
on $\rho_+$.
Thus the size of the bulk is not fixed by flux conservation.
To discuss the modulus dynamics, we take the moduli approximation,
namely assuming that $\rho_+\to \rho_+(x^{\mu})$.
Integrating over the extra dimensions, we obtain
\begin{eqnarray}
 \big(S_6\big)_{\rm mod}
=\frac{\pi\kappa M_6^4}{g^2}
 \int d^4 \tilde x 
 \left(-
  \frac{(\partial \rho_+)^2}{\rho_+}
  \right)
  \,.
\end{eqnarray}
After redefining the modulus as
\begin{eqnarray}
 \chi_6(x^{\mu})=
\sqrt{\frac{8\pi\kappa M_{6}^4}{g^2}\rho_{+}}
\,,\label{effrad6d}
\end{eqnarray}
we obtain the canonical form of the modulus kinetic term as
\begin{eqnarray}
 \big( S_6 \big)_{\rm mod}
= \int d^4 \tilde{x}
 \Big(
 -\frac{1}{2}(\tilde \partial\chi_6)^2
 \Big)\,.
\end{eqnarray}

\section{The one-loop effective potential of the volume modulus}

Next, we introduce a massless, minimally coupled scalar field and 
work in the Euclideanized space.
\footnote{For an earlier work on the Casimir effect in 6D spacetime,
see e.g., \cite{Milton}.}
The action for the massless scalar field perturbations is given by
\begin{eqnarray}
S_{\rm scalar}=\frac{1}{2}
                \int d^6 x\sqrt{g}
                \phi\Delta_6\phi\,. 
\label{scalaraction}
\end{eqnarray}

\subsection{Scalar one-loop effective action}

The one-loop effective action for a massless minimally coupled scalar field is
given by
\begin{eqnarray}
W_6= \frac{1}{2}{\rm ln }\,{\rm det} (-\Delta_6)\,,\label{effacc}
\end{eqnarray}
where $\Delta_6$ is 6D Laplacian.
$W_6$
needs to be regularized and renormalized. For this purpose, we define
\begin{eqnarray}
   W_s
 =-\frac{\mu^{2s}}{2}
 \int^{\infty}_0 \frac{dt}{t^{1-s}}
  {\rm Tr}\left(e^{-t(-\Delta_6)}\right)
\,,
\end{eqnarray}
where ${\rm Tr}\left(e^{-t(-\Delta_6)}\right)$ corresponds to
the (integrated) heat kernel.
The (integrated) zeta function is related to the heat kernel by 
a Mellin transformation:
\begin{eqnarray}
 \zeta(s,\Delta_6)
=\frac{1}{\Gamma(s)}
\int^{\infty}_0 dt\,
t^{s-1} {\rm Tr}\left(e^{-t(-\Delta_6)}\right)
= {\rm Tr} \Big((-\Delta_6)^s \Big)\,,
\label{zetafunc}
\end{eqnarray} 
and after analytically continuing to $s\to 0$ we obtain the renormalized one-loop effective action. 
The renormalized scalar field effective action can be written as
\begin{eqnarray}
W_{6, \rm ren}=-\frac{1}{2}\zeta'(0,\Delta_6)
              -\frac{1}{2}\zeta(0,\Delta_6)\ln \mu^2\,.
\end{eqnarray}
By integrating over the internal dimensions, the
4D effective potential is 
\begin{eqnarray}
W_{6,\rm ren}=\int \big(d^4 x \rho_+^2 \big) 
              V_{\rm 6, eff}
              =\int d^4{ \tilde x } 
              V_{\rm 6, eff}
               \,,\label{effact6d}
\end{eqnarray}
where $V_{\rm eff}$ has the dimensions $({\rm length})^{-4}$. 
For brevity, from now on we shall omit the subscript ``ren".

The zeta function is given by the summation
\begin{eqnarray}
\zeta(s,\Delta_6)
           =\int d^4 x 
            \sum_{m,n}
           \int \frac{d^4k}{(2\pi)^4}
           \frac{1}{\lambda^{2s}}\,,
\label{zeta6d}
\end{eqnarray}
where the eigenvalues are defined by 
\begin{eqnarray}
 \Delta_6 \phi_{\lambda}
=-\lambda^2 \phi_{\lambda}\,.
\end{eqnarray}
It is straightforward to show that 
\begin{eqnarray}
 \zeta(0,\Delta_6)
 = a_6(f=1)\,,
\end{eqnarray}
where $a_6(f)$ is a heat kernel coefficient,
defined by the asymptotic expansion of the heat kernel
\cite{VSV, Hoover}:
\begin{eqnarray}
{\rm Tr}\left(e^{-t(-\Delta_6)}\right)
\simeq 
\sum_{k\geq 0}
   t^{(k-6)/2}a_k(f)\,,
   \quad t\to 0\,.
\end{eqnarray}

\subsection{Continuous conformal transformations}

One strategy to evaluate the one-loop effective action 
and the effective potential is to define a continuous conformal transformation (parameterized by $\epsilon$)
\begin{eqnarray}
d{\tilde s}_{6,\epsilon}^2
 = e^{2(\epsilon-1)\omega}ds_6^2\,,\qquad
\omega= \frac{1}{2} \ln(2\rho)\,,
\label{conformal}
\end{eqnarray}
and thus
\begin{eqnarray}
 d{\tilde s}_6^2
=(2\rho)^{\epsilon}
 \left(
   \frac{dz^2}{g^2(1+z^2)^2}
 +\frac{g^2(1 - \alpha^2)^2 z^2}
          {4(1 + \alpha^2 z^2)^2}
    d\theta^2
 + d{\bf x}^2
 \right)\,,
\label{conf}
\end{eqnarray}
where for $\epsilon=1$ we have the original metric, which we shall denote as
$\Delta_{6,\epsilon}=\Delta_6$. The classical action of this scalar field
is changed under a conformal transformation Eq. (\ref{conformal})
\begin{eqnarray}
  S_{\rm scalar}
   =-\frac{1}{2}\int d^6 x\sqrt{g}
     \phi\Delta_6\phi
   =-\frac{1}{2}\int d^6 x\sqrt{\tilde g}
     \tilde \phi\Big(\tilde \Delta_6+E_6(\epsilon) \Big)
     \tilde \phi\,,
     \label{class6d}
\end{eqnarray}
where
\begin{eqnarray}
E_6(\epsilon)
&=& -4(\epsilon-1)^2 \tilde{g}^{ab}
                 \nabla_{a} \omega
                  \nabla_{b}\omega
     +2(\epsilon-1){\tilde \Delta}_6\ln\omega  
   \nonumber\\
 &=&
\left(\frac{1}{2\rho}\right)^{\epsilon}
\frac{g^2(1-\epsilon)(1 - \alpha^2) 
   \big\{
     (2+(1-\epsilon)z^2)
+  \alpha^2 z^2(-1+\epsilon-2z^2)
     \big\} }
     {(1 + \alpha^2 z^2)^2}\,.
     \label{good}
\end{eqnarray}
We perform zeta function regularization explicitly because
in the unwarped frame $\epsilon=0$ we can derive exact mass spectrum and thus
perform exact summing up of all the relevant KK modes (See Sec. IV). 
The correction associated with such a conformal transformation is commonly known as the cocycle function:
\begin{eqnarray}
  W_6
  &=&  
  -\frac{1}{2}
   \zeta'(0,\Delta_6)
  -\frac{1}{2}
   \zeta(0,\Delta_6)
   \ln \mu^2
\nonumber \\
& =& -\frac{1}{2}
   \zeta(0,\Delta_{6,\epsilon=0})
   \ln \mu^2
 +\Big\{
  -\frac{1}{2}
   \zeta'(0,\Delta_{6,\epsilon=0})
  -\int_0^1
    d\epsilon\,
    a_6\, (f=\partial_{\epsilon}\ln \Omega_{\epsilon})
\Big\}
\,.
\end{eqnarray}
The term $a_6$ is given by the volume integration of linear combinations of cubic order curvature invariants \cite{VSV,Hoover}:
\begin{eqnarray}
  a_6(f)
&:=& (4\pi)^{-3}\Bigl\{
\int_M d^6x \sqrt{g}\,
 \Big[ \frac f{7!}\Big(
      18R^{;A}{}_{;A}{}^{;B}{}_{;B}
    +17R_{;A}R^{;A}
    -2R_{AB;C}R^{AB;C} 
    -4R_{AB;C}R^{AC;B}
    +9R_{ABCD;E}R^{ABCD;E}
    \nonumber \\
&&\quad
    +28RR_{;A}{}^{;A}\
    -8R^{AB}R_{AB;C}{}^{;C} 
    +24R^{AB}R_{A}{}^{C}{}_{;BC}
    +12R_{ABCD}R^{ABCD ;E}{}_{;E}
    +\frac{35}{9}R^{3}
    -\frac{14}{3}RR_{AB}R^{AB} 
       \nonumber \\
&&\quad
    +\frac{14}{3}R R_{ABCD}R^{ABCD}
    -\frac{208}{9}R_{AB}R^{A}{}_{C}R^{BC}
       +\frac{64}{3}R_{AB}R_{CD}R^{AC BD}
      -\frac{16}{3}R_{AB}R^{A}{}_{CDE}R^{BCDE}
       \nonumber \\
&&\quad
   +\frac{44}{9}R_{ABCD}R^{AB}{}_{EF}R^{CD EF}   
  +\frac{80}{9}R_{ABCD}R^{A}{}_{E}{}^{C}{}_{F}
          R^{BEDF}
   \Big)
 \nonumber \\
&&\quad
     +\frac{f}{360} 
     \Big(
      6E_6{}^{;A}{}_{;A}{}^{;B}{}_{;B}
     +60E_6E_{6}{}^{;A}{}_{;A}
     +30E_{6}^{;A}E_{6;A}
     +60E_{6}^{3}
     +10R E_{6}{}^{;A}{}_{;A}
     +4R^{AB}E_{6;AB}
     +12R^{;A}E_{6;A}
     \nonumber \\
&&\quad
     +30E_6^2R
     +12E_6R^{;A}{}_{;A}
     +5E_6R^2
     -2E_6R^{AB}R_{AB}
     +2E_6R^{ABCD}R_{ABCD}
     \Big)   
 \Big]
     \nonumber\\
 &&
 \qquad
 +({\rm contribution\,\, of\,\, conical\,\, branes})
 \Big\}
   \,.\label{a6}
\end{eqnarray}

It is rather useful to use
the effective potential, as given by Eq. (\ref{effact6d}).
\begin{eqnarray}
V_{\rm 6, eff}(\alpha,\kappa,g,\mu;\rho_+)
=\frac{A_6(\alpha,\kappa,g)
      -B_6(\alpha,\kappa,g) 
     \ln(\mu^2 \rho_+)}{\rho_+^2}\,,
\label{originaleffpot}
\end{eqnarray}
where we define
\begin{eqnarray}
 \int d^4 x A_6(\alpha,\kappa,g)
 &=& \int d^4 \tilde x 
  \frac{ A_6(\alpha,\kappa,g)}{\rho_+^2}
  = -\int_0^1 
       d\epsilon 
       a_6(f=\frac{1}{2}\ln(\frac{2\rho}{\rho_+})) 
  -\frac{1}{2}\zeta'(0,\Delta_{6,\epsilon=0})
    \,,      \nonumber\\
\int d^4 x B_6(\alpha,\kappa,g)
 &=&\int d^4 \tilde x \frac{B_6(\alpha,\kappa,g)}{\rho_+^2}
  =\frac{1}{2}\zeta(0,\Delta_{6,\epsilon=0})\,.
 \label{6dcoef}
\end{eqnarray}
 Clearly, if $B_6(\alpha,\kappa,g)>0$,
 then the modulus effective potential has a minimum at
\begin{eqnarray}
 \rho_{+}^\ast
=\mu^{-2}  e^{(2A_6+ B_6)/(2B_6)}\,.\label{sta-san}
\end{eqnarray}
After a redefinition of the modulus, as given by Eq. (\ref{effrad6d}),
the effective potential can be rewritten as
\begin{eqnarray}
 V_{\rm 6,eff}(\alpha,\kappa,g,\mu;\rho_+):
=\Big(
\frac{8\pi\kappa M_6^4}{g^2}
 \Big)^2
\frac{A_6(\alpha,\kappa,g)
     -B_6(\alpha,\kappa,g)
       \ln
       \left(
       \frac{\mu^2 g^2\chi_6^2}
            {8\pi M_6^4\kappa}
       \right)}
     {\chi_6^4}\,. 
\end{eqnarray}
The field value at the minimum is then given by
\begin{eqnarray}
\chi_{6, \ast}^2
=\frac{8\pi\kappa M_6^4}
      {\mu^2 g^2}
e^{(2A_6+B_6)/(2B_6)}\,.\label{effmodu}
\end{eqnarray}


\subsection{Phenomenological implications after volume stabilization}

\subsubsection{For the hierarchy problem}

One of the most longstanding problems in phenomenology
is the hierarchy problem, namely
why gravity is so weak in comparison to the electroweak interaction
(why the Planck scale $M_{\rm Pl}\sim 10^{19}{\rm  GeV}$
is much larger than that of the electroweak scale $M_{\rm EW} \sim 10^{3} {\rm GeV}$).
A way to resolve the hierarchy problem in braneworld set-up
was first proposed in the large extra dimension model given in
Ref. \cite{Arkani-Hamed:1998rs}. The basic idea is that we assume that 
the fundamental gravitational scale is not $M_{\rm Pl}$ itself but the
higher-dimensional one ($M_6$ in 6D braneworld) and $M_{6}\sim M_{\rm EW}$.
Then, the observed Planck scale is effectively given as a result of a
dimensional reduction and in the present model
 \begin{eqnarray}
   M_{\rm pl}^2
 \simeq
 \frac{\rho_+(2\pi\kappa)}{g^2}
  M_{6}^4\,.     
 \end{eqnarray}
To get the observed value of reduced Planck scale,
the size of extra dimension should be $\big(\rho_+\big)^{1/2}\sim 0.1 {\rm mm}$. Thus now we ask whether volume stabilization at this scale is available
in the present model.

 If we assume a brane localized field, e.g., a Higgs field,
\footnote{It is difficult to treat any matter on a strictly
codimension two brane, unless we regularize the brane.
Thus, we implicitly assume that the brane has a small, 
but finite thickness rather than existing as a strictly codimension two defect.}
 whose bare mass is given by $m^2$
 on either brane at $\rho_{\pm}$ then the observed mass scales are
 \begin{eqnarray}
 m_{+}^2= m^2\,,\qquad 
 m_{-}^2= \alpha^2 m^2\,.
 \end{eqnarray}
We now assume that $m_{\pm}^2\sim M_{\rm EW}^2$.
Thus, the mass ratio between the field and the effective Planck mass is given by
\begin{eqnarray}
\frac{m_{+}^2}{M_{\rm pl}^2}
\simeq
\Big(  \frac{\mu^2 m^2}{M_{6}^4}\Big)
\Big(  \frac{g^2}{2\pi\kappa}\Big)
   e^{-(2A_6 +  B_6)/(2B_6)}\,,
   \qquad\qquad
 \frac{m_{-}^2}{M_{\rm pl}^2}
\simeq
\Big(  \frac{\mu^2 m^2}{M_{6}^4}\Big)
\Big(  \frac{g^2\alpha^2}{2\pi\kappa}\Big)
   e^{-(2A_6+B_6)/(2B_6)}\,.
  \end{eqnarray}
Assuming that the factor of $(\mu m/M_{6}^2)^2$ takes the optimal value of
${\cal O}(1)$ for the unification of all the fundamental energy scales in 6D,
the effective mass ratio is characterized by
\begin{eqnarray}
R(\alpha,\kappa,g):= \frac{g^2}{2\pi\kappa}
   e^{-(2A_6 + B_6)/(2B_6)}\Big|_{\rho_+=\rho_{+,\ast}}
\,, \label{ratio}
\end{eqnarray}
where we have used the value of $\rho_{+,\ast}$, given by Eq. (\ref{sta-san}).
As is explained above, 
once the size of the internal space $\rho_+^{1/2}$ is stabilized at $0.1{\rm mm}$, then $R$ has a value as $\sim 10^{-32}$
and our main task is to explore such a possibility 
in the parameter space $(\alpha,\kappa,g)$.


\subsubsection{For the cosmological constant problem}

Another important problem is the cosmological constant problem,
namely
why the energy density of the present dark energy,
assuming that its origin is vacuum energy of quantum fields,
$\rho_{\rm Vac}\sim (10^{-3}{\rm eV})^4$ is
much smaller than $M_{\rm Pl}^4$, expected from the standard model.
A characteristic property of 6D braneworld is that
the tree level vacuum energy of the brane, i.e., brane tension,
only changes the bulk deficit angle and does not affect the brane 
geometry \cite{Carroll:2003db}.
Note that there are several criticisms for such a self-tuning mechanism
mainly because flux conservation/quantization may
induce a hidden fine-tuning of brane tension \cite{cc2}.
Here we focus on how important the Casimir effect is 
at one-loop order of the vacuum energy, 
assuming that a self-tuning exists.
Of course, to build more realistic models with such a self-tuing mechanism 
is an important subject, but is out of scope of this article.

After volume stabilization,
from Eqs. (\ref{originaleffpot}) and (\ref{sta-san}),
the effective potential (Casimir energy density) takes the value
\begin{eqnarray}
V_{\rm 6, eff}^{\ast}(\alpha,\kappa,g)
=-\frac{1}{2}
         \mu^4 B_{6}(\alpha,\kappa,g)  e^{-(2A_6+B_6)/ B_6}
\Big|_{\rho_+=\rho_{+,\ast}} 
(=\rho_{\rm Cas})
\,,
\label{vacuum}
\end{eqnarray} 
and hence, the realized brane vacuum energy is almost completely determined by the renormalization.
The renormalization scale $\mu$ is also a free parameter but
in order for an additional hierarchy problem not to appear
it should be set to a value such as $\mu\sim M_6 (\sim M_{EW} )$. 
Then, $B_6\propto g^4$ and thus we get the relation depicted Eq. (\ref{ever})
with $n=2$.
Note that Casimir energy density becomes negative as in Eq. (\ref{vacuum}). 
Thus, some additional mechanisms to uplift the potential minimum must exist.
In this article, however, we just focus on the absolute value of the energy density.


\section{Evaluation of the derivative of zeta function in the unwarped frame}

\subsection{Mass spectrum}

From Eq. (\ref{class6d}), the mass spectrum in the unwarped frame
is determined by 
\begin{eqnarray}
 \left(
\tilde   \Delta_6
+ E_6(0)
  \right)
  \tilde \phi_{\lambda} =
-\lambda^2\tilde
\phi_{\lambda}\,.
\label{eigeneq}
\end{eqnarray}
We shall decompose the mass eigenfunction as
\begin{eqnarray}
 \tilde \phi_{\lambda}
=\int
  \frac{d^4k}{(2\pi)^2}
  \sum_{m,n}
  \Phi_{\lambda}(z)e^{in \varphi} e^{i {\bf kx }}\,.
\end{eqnarray}
The equation of motion of equation (\ref{eigeneq}) has a series of exact solutions \cite{Carter:2006uk, Parameswaran:2006db,emn}:
\begin{eqnarray}
\Phi_{\lambda}(z)&=&
\sqrt{\frac{1+\alpha^2 z^2}{1+z^2}}
\Big[
 A \left(\frac{z^2}{1+z^2}\right)^{-n /2\kappa}
   \left(\frac{1}{1+z^2}\right)^{ n \alpha^2/2\kappa}
{}_{2}F_{1}\big(1-\nu-\frac{n}{2\kappa} (1-\alpha^2),
            \nu-\frac{n}{2\kappa} (1-\alpha^2),
            1- \frac{n}{\kappa};
           \frac{z^2}{1+z^2}
           \big)
\nonumber\\
&+& B
   \left(\frac{z^2}{1+z^2}\right)^{n/2\kappa}
   \left(\frac{1}{1+z^2}\right)^{-n \alpha^2 /2\kappa}
{}_{2}F_{1}\big(
           1-\nu+\frac{n}{2\kappa} (1-\alpha^2),
            \nu+\frac{n}{2\kappa} (1-\alpha^2),
            1+\frac{n}{\kappa} ;
           \frac{z^2}{1+z^2}
           \big) \label{nonzero}
\Big],
\end{eqnarray}
where ${}_{2}F_{1}(a,b,c;x)$ is Gauss's hypergeometric function and
\begin{eqnarray}
\nu=\frac{1}{2}
 \left(
  1+\sqrt{1+\frac{\lambda^2-k^2}{g^2} +\frac{n^2}{\kappa^2}(1-\alpha^2)^2}
 \right).
\end{eqnarray}
Here imposing the regularities on both conical branes, we obtain the following
exact mass spectrum
\begin{eqnarray}
\lambda^2= k^2 +g^2 \big[4m(m+1)+\frac{2|n|}{\kappa}(2m+1)(1+\alpha^2)
          +\frac{4n^2\alpha^2}{\kappa^2}\big],\qquad\qquad
m=0,1,2,\cdots.
\label{eigen}
\end{eqnarray}
Our method to determine the mass spectrum is essentially based
on the same arguments given in \cite{Parameswaran:2006db}
(more precisely, it is demanded that the wave functions should be 
Hermitian) . However, if only normalizability were imposed this
would allow for logarithmic divergences at the poles and this may well lead to additional modes in the eigenvalue spectrum.
But what should be stressed
is that as is discussed in the case of 4D toy model \cite{emn}
the effects of these additional modes are not important
and neglecting them is valid in the following discussions.

\subsection{Zeta function regularization}

The first step
 is the integration of the zeta function Eq. (\ref{zeta6d})
over the Fourier space
\begin{eqnarray}
(2\pi)^4\zeta (s,\Delta_{6,\epsilon=0})= \int d^4x
  \frac{\pi^2 g^{2(2-s)}}{(s-1)(s-2)}
\sum_{m,n}
 \big[4m(m+1)+\frac{2|n|}{\kappa}(2m+1)(1+\alpha^2)  
 +\frac{4n^2\alpha^2}{\kappa^2}\big]^{2-s}
\,.
\label{zetafunk}
\end{eqnarray}
Hereafter, we omit the index $\Delta_{6,\epsilon=0}$.
For a later convenience, we define
\begin{eqnarray}
a=4, \qquad\qquad b= {4(1+\alpha^2)\over \kappa},\qquad\qquad 
c={4\alpha^2\over \kappa^2}, \qquad\qquad
q=-1,\qquad\qquad \beta=1/2\,,
\end{eqnarray}
and using them
\begin{eqnarray}
&&\hat \beta(n) =\beta+\frac{b n}{2a}
              =\frac{1}{2}+\frac{(1+\alpha^2)|n|}{2\kappa},
\nonumber\\
&&\hat q(n) =q + c n^2-\frac{(bn)^2}{4a}
         = -\frac{n^2}{\kappa^2} (1-\alpha^2)^2 -1
\,. \label{nsum}
\end{eqnarray}

\subsection{Extended binomial expansion}

Frequently, the form of the two-dimensional Epstein zeta function
allows one to perform its summation in an elegant way which involves
the Chowla-Selberg expansion formula or more frequently generalization
 of it, see \cite{mns, emn, ElizaldeCS}.
Some conditions must be satisfied;
the most impotant one is that
the quadratic form must be positive definite and the
constant $q$ term should be also non-negative. But this is not here
the case, and the fact that
  $q<0$ does not allow for such a beautiful analysis.
 Henceforth, we shall apply in what follows what we have called the
{\it extended binomial expansion} approach, namely, we
will introduce an extra summation via
\begin{eqnarray}
&& \sum_{m=0}^{\infty}
   \sum_{n=1}^{\infty}
 \big[a(m+\beta)^2+b(m+\beta)n+c n^2+q \big]^{-s+2}
 \nonumber\\
&=&\sum_{m=0}^{\infty}
   \sum_{n=1}^{\infty}
 \big[a(m+\hat \beta)^2+\hat q \big]^{-s+2}
=\sum_{m=0}^{\infty}
   \sum_{n=1}^{\infty}
   \sum_{j=0}^{\infty}
  \frac{\Gamma(3-s)}
       {\Gamma(3-s-j)j!}
  \big[a(m+\hat \beta)^2\big]^{2-s-j}
  {\hat q}^{j}
\nonumber \\
&=&\sum_{m=0}^{\infty}
   \sum_{n=1}^{\infty}
   \sum_{j=0}^{\infty}
  \frac{(-1)^j\Gamma(s+j-2)}
       {\Gamma(s-2)j!}
  \big[a(m+\hat \beta)^2\big]^{2-s-j}
  {\hat q}^{j}
\,,
\end{eqnarray}
where the validity of the binomial expansion is defined for
\begin{eqnarray}
  \Big|
\frac{\hat q}{a(m+\beta)^2}
  \Big|<1\,,
\end{eqnarray}
which is indeed satisfied for all possible values of $m$ and $n$ in
our model. 
Until now, we can reduce to the following expression of the zeta
function
\begin{eqnarray}
&&(2\pi)^4\zeta(s)
\nonumber\\
&=&
\int d^4 x
 \sum_{j=0}^{\infty}
 \sum_{m,n}
\Big(\frac{2^{4-2s-2j}
\pi^2 g^{2(2-s)}\Gamma(s+j-2)}{j!\Gamma(s)}\Big)  
  \Big[m+\frac{1}{2} +\frac{1+\alpha^2}{2\kappa}\Big]^{2(2-s-j)}
  \Big[\frac{n^2}{\kappa^2}(1-\alpha^2)^2 +1\Big]^j\,.
\end{eqnarray}
The j-summation is absolutely converging and so we can exchage
the order of $j$-summation and $(m,n)$-summation.
Then, to perform $m$-summation is rather straightforward.
We decompose the zeta function into the contributions of non-axisymmetric
and axisymmetric modes as
\begin{eqnarray}
\zeta(s)=\zeta(s)\Big|_{n\neq 0}+\zeta_{0}(s)\,.
\end{eqnarray}
The contribution of axisymmetric modes is discussed in Appendix B and
hereafter we focus on non-axisymmetric modes with $n\neq 0$.
Then, we get 
\begin{eqnarray}
&&(2\pi)^4\zeta(s)\Big|_{n\neq 0}
\nonumber\\
&=&
\int d^4 x
 \sum_{j=0}^{\infty}
 \sum_{n=1}^{\infty}
\Big(\frac{2^{5-2s-2j} \pi^2 g^{2(2-s)}\Gamma(s+j-2)}{j!\Gamma(s)}\Big)  
  \Big[\frac{n^2}{\kappa^2}(1-\alpha^2)^2 +1\Big]^j
\zeta_{H}\big(2s+2j-4,\frac{1}{2}+\frac{1+\alpha^2}{2\kappa} n\big)\,.
\end{eqnarray}
The zeta function is convergent only ${\rm Re}(s)>5/2$
and thus we need to analytic continuation to $s\to 0$.

\subsection{Zeta function regularization}

By subtracting the divergent terms from the asymptotic expansion of 
the zeta function (in the limit of $s\to 0$) and adding back the 
counterterms we obtain
\begin{eqnarray}
(2\pi)^4 \zeta(s)\Big|_{n\neq 0}
&=&\int d^4 x
\left(
  P(s)
+
\sum_{j=0}^{\infty}
 \Delta(j,s)
 G(j,s)
\right)
\,,
\label{nonzeron}
\end{eqnarray}
where
\begin{eqnarray}
P(s):= \sum_{j=0}^{\infty}
   G(j,s) \Big[
   \sum_{n=1}^{\infty}
     \Big\{
    \big[\frac{n^2}{\kappa^2})(1-\alpha^2)^2 +1\big]^j
    \zeta_H \big(2s+2j-4,\frac{1}{2}+\frac{1+\alpha^2}{2\kappa}n\big)
    -F(n,j,s)
     \Big\}
   \Big]\,,\label{P}
\end{eqnarray}
and
\begin{eqnarray}
G(j,s)&=&\frac{2^{5-2j-2s} \Gamma(s+j-2)}{j!\Gamma(s)}
        \pi^2 g^{2(2-s)}\,,
\nonumber\\
 F(n,j,s)
&:=&\Big(\frac{n}{\kappa}\Big)^{-1-2s}
\frac{2^{-5+2s+2j}(1-\alpha^2)^{2j}}
     {(2s+2j-5)(1+\alpha^2)^{2s+2j-5}}
\nonumber\\
&\times&
\Big\{ w_0 (\alpha,j,s) \big(\frac{n}{\kappa}\big)^6
+w_1 (\alpha,j,s) \big(\frac{n}{\kappa}\big)^4
+w_2 (\alpha,j,s) \big(\frac{n}{\kappa}\big)^2
+w_3 (\alpha,j,s) 
\Big\}\,,
\nonumber\\
\Delta(j,s)
&=&\frac{2^{-5+2s+2j} \kappa^{2s+1}(1+\alpha^2)^{5-2s-2j}(1-\alpha^2)^{2j}}
     {2s+2j-5}
\nonumber\\
&\times&
\Big\{
\frac{w_0 (\alpha,j,s)}{\kappa^6}  \zeta_R(2s-5)
+\frac{w_1 (\alpha,j,s)}{\kappa^4} \zeta_R(2s-3)
+\frac{w_2 (\alpha,j,s)}{\kappa^2} \zeta_R(2s-1)
\nonumber \\
&+&w_3 (\alpha,j,s) \zeta_R(2s+1)
\Big\}\,.
\end{eqnarray}


\subsection{Derivative of zeta functions}

\subsubsection{Analytic continuation of the deriavative of $\zeta\,-$~function}

We first make an analytic continuation of the subtracted zeta function.
The derivative of the subtracted zeta function  $P'(0)$ is given in the Appendix A in detail. We show a typical example of this function in Fig.~1.
The $j$ and $n$-summations both show good convergency.

\subsubsection{Analytic continuation of counter terms}

The analytic continuation of the counterterms are given by
\begin{eqnarray}
&&\frac{d}{ds}\sum_{j=0}^{\infty}\Big(G(j,s)\Delta(j,s)\Big)
\nonumber\\
&=&\frac{d}{ds}\Big(G(0,s)\Delta(0,s)\Big)\Big|_{s\to 0}
+\frac{d}{ds}\Big(G(1,s)\Delta(1,s)\Big)\Big|_{s\to 0}
+\frac{d}{ds}\Big(G(2,s)\Delta(2,s)\Big)\Big|_{s \to 0}
+\sum_{j=3}^{\infty}
 \frac{d}{ds}\Big(G(j,s)\Delta(j,s)\Big)\Big|_{s \to 0}\,.
\nonumber\\
&&\label{GD6D}
\end{eqnarray}
The result of analytic continuation of each $j$ term is shown in Appendix A.
We show an example showing the convergency of the summation in Fig. 2.

\begin{figure}
\begin{center}
  \begin{minipage}[t]{.45\textwidth}
   \begin{center}
    \includegraphics[scale=.8]{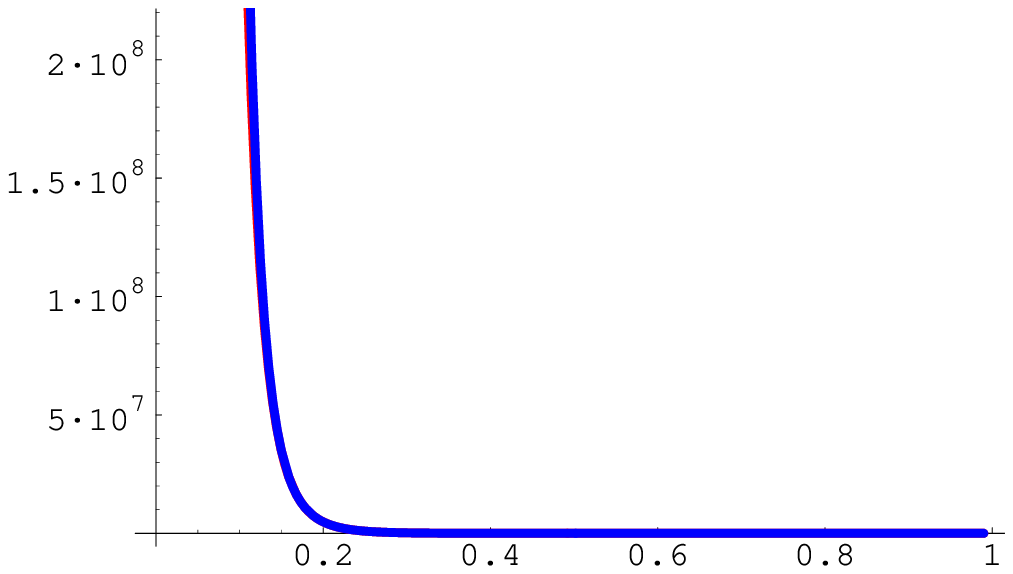}
        \caption{Plots for $ P'(0)$, Eq. (\ref{PD6D}), as a function of $r$ are shown for $\kappa=0.9$.
The red and blue curves correspond to truncation of the $j$-summation at $j_{\rm max}=100$ and $j_{\rm max}=200$, respectively.
       }  
   \end{center}
   \end{minipage} 
\hspace{0.5cm}
   \begin{minipage}[t]{.45\textwidth}
   \begin{center}
    \includegraphics[scale=.8]{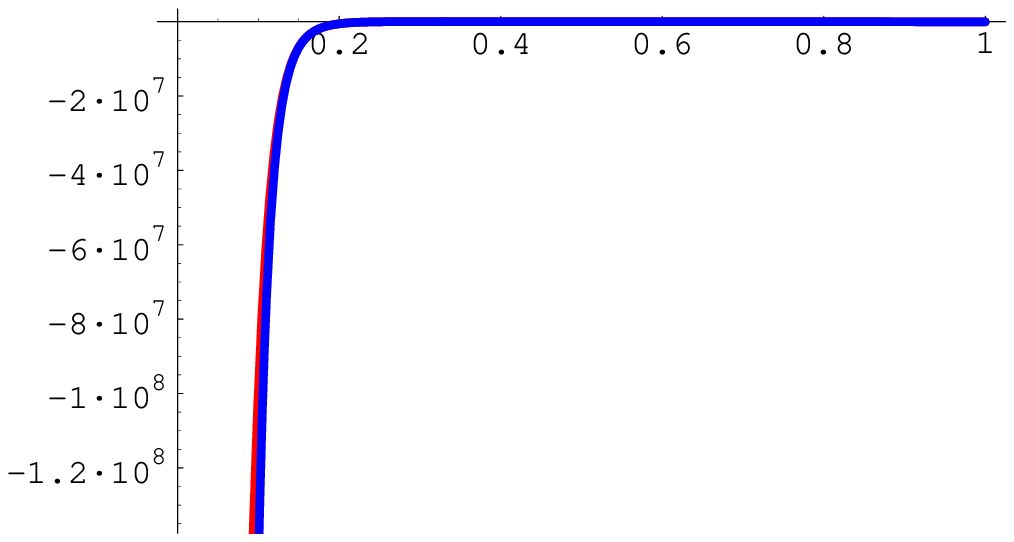}
\caption{Plots for $(d/ds) (\sum_j G\Delta)(s)|_{s\to 0}$, Eq. (\ref{GD6D}), as a function of $\alpha$ are shown for $\kappa=0.9$.
 The red and blue curves correspond to truncation of $j$-summation at $j_{\rm max}=100$ and $j_{\rm max}=200$, respectively.}          
   \end{center}
   \end{minipage}
   \end{center}
\end{figure}


\subsection{On the heat kernel coefficients and cocycle functions}

As is explained in Sec. III, the coefficient $B_6(\alpha,\kappa,g)$
in the one-loop effective potential can be obtained by
\begin{eqnarray}
\int d^4 x B_6(\alpha,\kappa,g)
 &=&\frac{1}{2}\zeta(0,\Delta_{6,\epsilon=0})\,. \label{evab} 
\end{eqnarray}
Here $B_6(\alpha,\kappa,g)$ includes both brane and bulk contributions.
To decompose $B_6(\alpha,\kappa,g)$ into brane and bulk contributions, 
it is useful to focus on the fact that
$a_6(f=1)=\zeta(0,\Delta_{6,\epsilon=0})$ \cite{VSV}.
The bulk heat kernel is given by
\begin{eqnarray}
 a_{6,{\rm bulk}}(f=1)
=\int d^4x
\frac{\kappa g^4}
     {40320\pi^2}
\int_0^{\infty}dz\,
\frac{z}{(1+z^2)(1+\alpha^2 z^2)}
\frac{{\cal F}(z)}{(1+\alpha^2z^2)^6}\,, \label{aa6}
\end{eqnarray}
where ${\cal F}(z)$ is a 12th order polynomial of $z$ given in Appendix C.
Note that now there is no way to evaluate the brane heat kernel
 $a_{6,{\rm brane}}(f=1)$ because of a lack of mathematical formulation
of conical heat kernel.
Thus the pure bulk and brane contributions are evaluated by
\begin{eqnarray}
&& B_{6,{\rm bulk}}(\alpha,\kappa,g)
=\frac{\kappa g^4}
     {80640\pi^2}
\int_0^{\infty}dz\,
\frac{z}{(1+z^2)(1+\alpha^2 z^2)}
\frac{{\cal F}(z)}{(1+\alpha^2z^2)^6} \,,
\label{b6}
\nonumber\\
&& B_{6,{\rm brane}}(\alpha,\kappa,g)
= B_{6}(\alpha,\kappa,g)- B_{6,{\rm bulk}}(\alpha,\kappa,g)\,,
\end{eqnarray}
respectively.
The authors of Ref. \cite{mns} studied the contribution of 
conical branes to the total heat kernel and they conclude
that these contributions are not negligible.
We will see them in the next section.


The cocycle function from the bulk part is given by
\begin{eqnarray}
&&-\int_0^1 d\epsilon a_6(f=\frac{1}{2}\ln\big(\frac{2\rho}{\rho_+}\big))
\nonumber\\
&=&
-\int d^4x \frac{\kappa g^4}{80640\pi^2}
  \int_0^1  d\epsilon
  \int_0^{\infty} dz\,
    \frac{z}{(1+z^2)(1+\alpha^2z^2)}
    \frac{{\cal G}(\epsilon, z)}
          {(1+\alpha^2 z^2)^6}
\ln\Big(2\sqrt{\frac{1+\alpha^2 z^2}
                    {1+z^2}}\Big)\,,
\label{bb6}
\end{eqnarray}
where ${\cal G}(\epsilon,z)$ is also a 12th order polynomial
given explicitly in Appendix C.
Thus, we also obtain the coefficient
$A_{6,{\rm bulk}}(\alpha,\kappa,g)$ as
\begin{eqnarray}
&&A_{6,{\rm bulk}}(\alpha,\kappa,g)
\nonumber\\
&=&-\int_0^1 d\epsilon
 \frac{\kappa g^4}{80640\pi^2}
   \int_0^{\infty} dz\,
    \frac{z}{(1+z^2)(1+\alpha^2z^2)}
    \frac{{\cal G}(\epsilon,z)}
          {(1+\alpha^2 z^2)^6}
\ln\Big(2\sqrt{\frac{1+\alpha^2 z^2}
                    {1+z^2}}\Big)\,.\label{a6}
\end{eqnarray}
Note that the total coefficient $A_6(\alpha,\kappa,g)$ is given by
\begin{eqnarray}
\int d^4x  A_6(\alpha,\kappa,g)
&=& -\frac{1}{2}\zeta'(0,\Delta_{6,\epsilon=0})
   +\int d^4x A_{6,{\rm bulk}}(\alpha,\kappa,g)
   +\int d^4x A_{6,{\rm brane}}(\alpha,\kappa,g)
\nonumber\\
&=&  \int d^4 x \tilde  A_6(\alpha,\kappa,g)
   + \int d^4 x A_{6,{\rm brane}}(\alpha,\kappa,g)\,,
\label{tildea}
\end{eqnarray}
where $A_{6,{\rm brane}}(\alpha,\kappa,g)$ is
the brane part of the cocycle function and $\tilde A_{6}
(\alpha,\kappa,g)$ is the total of the 1st and 2nd term
of the middle step, which now we can evaluate.
As the case of $B_{6,{\rm brane}}(\alpha,\kappa,g)$,
 we have no direct way 
to evaluate $A_{6,{\rm brane}}(\alpha,\kappa,g)$.
In the following evaluations, we will use
$\tilde A_{6}(\alpha,\kappa,g)$
instead of $A_{6}(\alpha,\kappa,g)$ to discuss phenomenological 
implications without the brane cocycle function.
The evaluation of $A_6(\alpha,\kappa,g)$ is left for a future work.
Note the brane contribution is also partially included in
$\zeta'(0,\Delta_{6,\epsilon=0})$ and thus
we believe that $A_{6,{\rm brane}}$ should not be dominant.


\section{Volume stabilization and Phenomenological implications}

\subsection{Volume stabilization}

In Fig. 3, we plot
$B_{6,{\rm bulk}}(\alpha, \kappa,g)$ (and thus the density of $a_6(f=1)$)
 given in Eq. (\ref{b6})
and the integrand of $\zeta(0,\Delta_{6,\epsilon=0})$ as functions of $\alpha$
for several choices of $\kappa$.
An important observation is that for smaller $\alpha$
the sign of the integrand of $\zeta(0,\Delta_{6,\epsilon=0})$
becomes negative
implying that the volume modulus is {\it destabilized}.
In the case of 4D toy model discussed in Ref \cite{mns},
there is also negative brane contruibution but then
the bulk effect still dominates and the modulus is always stabilized.
For any value of $\kappa$, we obtain the critical value of $\alpha_{\ast}(\kappa)$, below which the volume modulus is destabilized.
In Fig. 4, we show the critical $\alpha_{\ast}$ as a function
of $\kappa$.

\begin{figure}
\begin{center}
  \begin{minipage}[t]{.45\textwidth}
   \begin{center}
    \includegraphics[scale=.8]{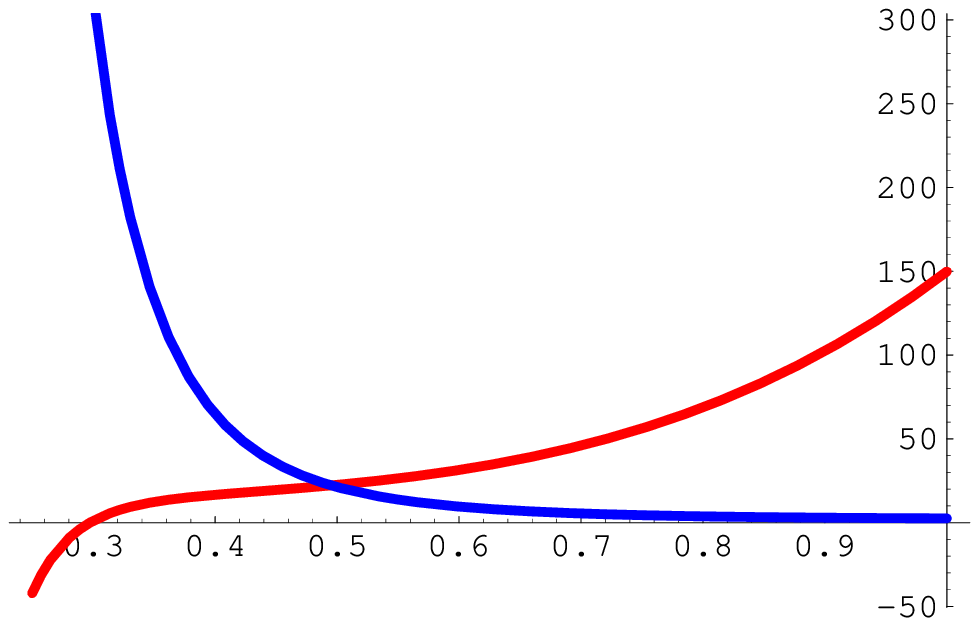}
  \caption{A plot for $B_6(\alpha,0.4,10)$ (the blue curve)
and the integrand
of $\zeta(0)$  for the same model parameters (the red curve)
as a function of $\alpha$ is shown.
We take $j_{max}=100$ and $n_{\max}=20$.}  
   \end{center}
   \end{minipage} 
\hspace{0.5cm}
   \begin{minipage}[t]{.45\textwidth}
   \begin{center}
    \includegraphics[scale=.8]{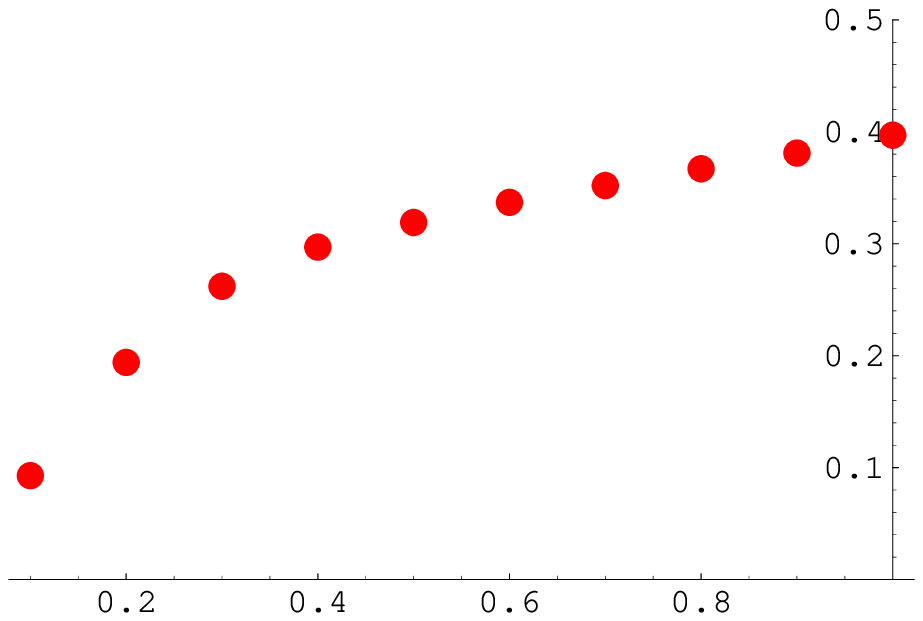}
\caption{A plot for critical warping $\alpha_{\ast}(\kappa)$
is shown as a function of $\kappa$.}          
   \end{center}
   \end{minipage}
   \end{center}
\end{figure}

\subsection{Implications for the hierarchy problem}

The resultant mass ratio of a brane localized field
to the effective Planck scale is characterized by Eq. (\ref{ratio}).
Our interest is the logarithmic scale of the ratio:
\begin{eqnarray}
\log_{10}\Big(R(\alpha,\kappa,g)\Big)
=\log_{10}\Big(\frac{g^2}{2\pi \kappa}\Big)
-\Big(
 \frac{\tilde A_6(\alpha,\kappa,g)}{B_6(\alpha,\kappa,g)}
+\frac{1}{2}
 \Big)\log_{10}e\,. \label{loghrc}
\end{eqnarray}
Note that we use $\tilde A_6$ in Eq. (\ref{tildea}) instead of $A_6$.
For the evaluation of the heat kernel coefficient
$B_6(\alpha,\kappa,g)$, we use Eq. (\ref{evab}).
As we have seen, there is critical values of $\alpha_{\ast}(\kappa)$
for each $\kappa$, below which the modulus is destabilized.
In Fig. 5, we show the plot of $\log_{10}(R(\alpha,\kappa=0.3,g))$
as a function of $\alpha>\alpha_{\ast}(0.3)$
for $g=10^{-3},10^{-1},10$.
A realistic value of $\log_{10} (R)\sim -32$ is possible,
but the warping parameter $\alpha$ should be close to $\alpha_{\ast}(\kappa)$
and thus, a fine-tuning is required though it is not as
severe as the original fine-tuning.


\subsection{Implications for the cosmological constant problem}

We can also discuss the implications for the cosmological constant problem
by evaluating the order of the effective energy density of the volume
modulus. In Eq. (\ref{vacuum}), we made the optimal choice of
 $\mu\sim M_6\sim m$.
We show the plot of
\begin{eqnarray}
\log_{10}\Big(
       \Big| 
       \frac{V^{\ast}{}_{6,{\rm eff}}(\alpha,\kappa,g)}{\mu^4} 
       \Big|
      \Big)
=\log_{10}\Big(\frac{B_6(\alpha,\kappa,g)}{2}\Big)
-\Big(
 2\frac{\tilde A_6(\alpha,\kappa,g)}
        {B_6(\alpha,\kappa,g)}
+1
 \Big)\log_{10}e\,.
\end{eqnarray} 
The degree of the ratio is somewhat sensitive to the value of
the bulk dilaton coupling especially for smaller values.
As is expected, the result is very similar to the case of the hierarchy problem.

It appears to be possible to obtain an observationally acceptable value of the
effective cosmological constant on the brane, but again
$\alpha$ should be close to the critical value
$\alpha_{\ast}(\kappa)$, as for the previous case.
In Fig. 6, we show the plot of
$\log_{10}\big(|V^{\ast}{}_{6,{\rm eff}}(\alpha,0.3,g)/\mu^4|\big)$
as a function of $\alpha>\alpha_{\ast}(0.3)$ for $g=10^{-3},10^{-1},10$.

\begin{figure}
\begin{center}
  \begin{minipage}[t]{.45\textwidth}
   \begin{center}
    \includegraphics[scale=.8]{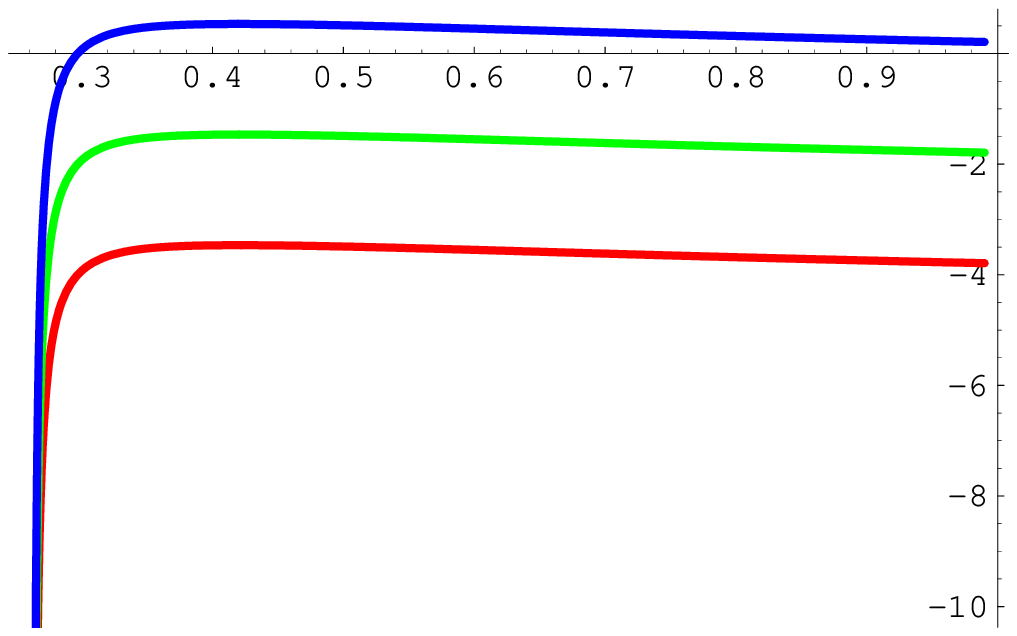}
        \caption{ $\log_{10}(R(\alpha,0.3,g))$ is shown as a function
of $\alpha$.
The red and blue curves correspond to the cases that
$g=10^{-3},10^{-1},10$, respectively.
Note that $\alpha>\alpha_{\ast}(\kappa=0.3)$.
We take $n_{\rm max}=20$ and $j_{\rm max}=100$.}
   \end{center}
   \end{minipage}
   \hspace{0.5cm}
   \begin{minipage}[t]{.45\textwidth}
   \begin{center}
    \includegraphics[scale=.8]{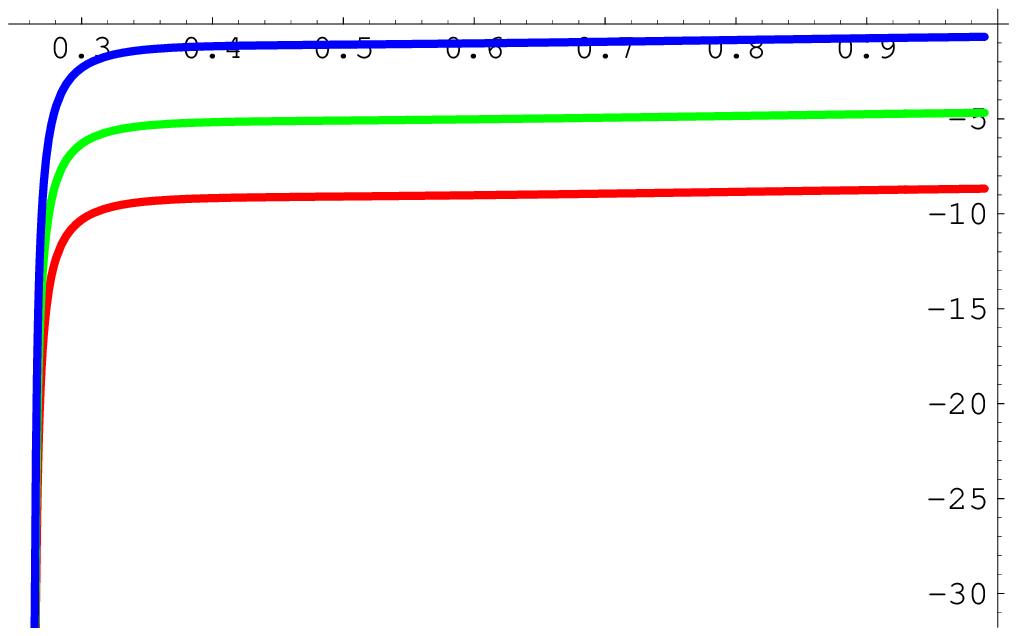}
\caption{$\log_{10}(
        |V^{\ast}{}_{6,{\rm eff}}(\alpha,0.3,g)/\mu^4| 
        )$ is shown as a function
of $\alpha$.
The red, green and blue curves correspond to the cases that
$g=10^{-3},10^{-1},10$, respectively.
Note that $\alpha>\alpha_{\ast}(\kappa=0.3)$.
We take $n_{\rm max}=20$ and $j_{\rm max}=100$.}
   \end{center}
   \end{minipage}
   \end{center}
\end{figure}


\section{Conclusion}

In this article, we have discussed the Casimir effect in a 6D warped flux compactification model based on a 6D supergravity
and its implications for phenomenology, i.e., the hierarchy and cosmological
constant problems. In its 4D effective theory of the model, 
a volume modulus appears and 
to stabilize the modulus we considered the Casimir effect induced by a
massless, minimally coupled bulk scalar field.
The effective potential of the volume modulus has the form
of a Coleman-Weinberg potential.
The stability itself can be determined by the sign of the 
coefficient in front of the logarithmic term $B_6(\alpha,\kappa,g)$.

There has been no mathematical formulation of the heat kernel
coefficient for 6D conical branes as far as the author is aware of.
However, we now have an exact mass spectrum in the unwarped frame 
and thanks to the fact that $a_6(f=1)=\zeta(0,\Delta_{\epsilon=0})$,
 by summing up all the modes with zeta function regularization,
 we can obtain the value of $B_6$.
As a result, especially for $\alpha<\alpha_{\ast}(\kappa)$,
where $\alpha_{\ast}(\kappa)$ is deficit angle dependent critical
value, the volume modulus is destabilized because of a strong negative
contribution from the brane quantum corrections.
In contrast, for $\alpha>\alpha_{\ast}(\kappa)$ the modulus is stabilized.
After volume stabilization, we then discussed the implications for the
hierarchy and cosmological constant problems and showed that it is possible to get observationally acceptable values of the ratio of effective mass scales and vacuum energy density.
However, for each value of the deficit angle of the reference $(+)$-brane,
the value of $\alpha$ should be tuned to be close to $\alpha_{\ast}$.

As is mentioned in the text, in the framework of the present model it is not possible to obtain a positive vacuum energy density and this fact 
 requires some modification of the present model, e.g., some kind of uplifting mechanism and/or field content contributions to the Casimir effect from the various multiplets that arise in the original supergravity model. In future work, we will explore more realistic modifications of the present model. 
Concerning the background model, we need to find a 6D braneworld model
where a self-tuning mechanism at tree level exists.
It would also be intersting to consider the Casimir effect in a cosmological
(time-dependent) background. We hope to report on these in future publications.

\section*{Acknowledgements}
This work was supported in part by the project ``Transregio (Dark
Universe)". The author would like to thank E.~Elizalde and W.~Naylor
for reading an earlier version of this manuscript
and making useful suggestions.

\begin{description}

\item{\it Note added:}

${\quad}$After completing this work, the author was informed
a related work Ref. \cite{Frank:2007jb}. 
In this work,
the Casimir force between two parallel plates
in Randall-Sundrum model \cite{Randall:1999ee}
was computed and bounds on the brane separation
relative to the bulk curvature radius
from Casimir force measurements were obtained,
whose value is relevant to the possibility of the resolution
of the hierarchy problem in this model.
\end{description}

\appendix

\section{Results of analytic continuations}

\subsection{Asymptotic expansion and analytically continued zeta function}

To perform analytic continuation, we derive the asymptotic expansion
(See e.g., \cite{ElizaldeBook})
\begin{eqnarray}
&&\zeta_{H}\big(2s+2j-4,\frac{1}{2}+\frac{1+\alpha^2}{2\kappa} n\big)
\nonumber\\
&=&
\frac{1}{2s+2j-5}\Big(\frac{1+\alpha^2}{2\kappa}\Big)^{5-2s-2j}
 \big(n+\frac{\kappa}{1+\alpha^2}\big)^{5-2s-2j}
+\frac{1}{2}\Big(\frac{1+\alpha^2}{2\kappa}\Big)^{4-2s-2j}
 \big(n+\frac{\kappa}{1+\alpha^2}\big)^{4-2s-2j}
\nonumber\\
&+&\frac{1}{2s+2j-5}
\sum_{k=2}^{\infty}
 \frac{B_k}{k!}
 (2s+2j-5)_k
   \Big(\frac{1+\alpha^2}{2\kappa}\Big)^{(5-k)-2s-2j}
   \big(n+\frac{\kappa}{1+\alpha^2}\big)^{(5-k)-2s-2j}
 \,,
\end{eqnarray}
where $B_k$ is the Bernoulli numbers.
Following the above fact, we obtain the desired asymptotic expansion
\begin{eqnarray}
&&\Big[\frac{n^2}{\kappa^2}(1-\alpha^2)^2 +1\Big]^j
\zeta_{H}\big(2s+2j-4,\frac{1}{2}+\frac{1+\alpha^2}{2\kappa} n\big)
\nonumber\\
&=&\frac{2^{5-2s-2j} (1+\alpha^2)^{5-2s-2j}(1-\alpha^2)^{2j}}
        {2s+2j-5}
\Big(\frac{n}{\kappa}\Big)^{-1-2\kappa}
\nonumber\\
&\times&
\Big(
 w_0 (\alpha,j,s) \big(\frac{n}{\kappa}\big)^6
+w_1 (\alpha,j,s) \big(\frac{n}{\kappa}\big)^4
+w_2 (\alpha,j,s) \big(\frac{n}{\kappa}\big)^2
+w_3 (\alpha,j,s) 
\Big)
\,.
\end{eqnarray}
where $w_i(\alpha,j,s) \,\, (i=0,\cdots,3)$ are given by 
\begin{eqnarray}
 w_0 (\alpha,j,s) &=&1\,,
\nonumber\\
 w_1 (\alpha,j,s) &=&\frac{-\left( 2\,j^2\,{\left( -1 + {\alpha }^2 \right) }^2 + \left( 10 - 9\,s + 2\,s^2 \right) \,{\left( -1 + {\alpha }^2 \right) }^2 + 
      4\,j\,\left( s\,{\left( -1 + {\alpha }^2 \right) }^2 - 3\,\left( 1 - {\alpha }^2 + {\alpha }^4 \right)  \right)  \right) }{3\,
    {\left( -1 + {\alpha }^2 \right) }^2\,{\left( 1 + {\alpha }^2 \right) }^2}
\,,
\nonumber\\
w_2 (\alpha,j,s)& =&
\Bigg(\big( -5 + 2\,j + 2\,s \big) \,
  \Big( -4\,\big( -2 + j + s \big) \,\big( -1 + j + s \big) \,
\big( -3 + 2\,j + 2\,s \big) 
\nonumber\\
&+& 30\,\big( -2 + j + s \big) 
 \,{\big( 1 + {\alpha }^2 \big) }^2\,
  \Big( \frac{j}{{\left( -1 + {\alpha }^2 \right) }^2} + 
         \frac{\left( -1 + j + s \right) \,\left( -3 + 2\,j + 2\,s \right)}
              {{\big( 1 + {\alpha }^2 \big) }^2} \Big)  
\nonumber\\
&-& 15\,\big( -2 + j + s \big) 
\,\Big( 2\,\big( -1 + j + s \big) 
        \,\big( -3 + 2\,j + 2\,s \big)  + 
         \frac{6\,j\,{\big( 1 + {\alpha }^2 \big) }^2}
     {{\big( -1 + {\alpha }^2 \big) }^2} \Big)
\nonumber\\
& +& \frac{15\,{\big( 1 + {\alpha }^2 \big) }^4
\,\Big( \frac{3\,\big( -1 + j \big) \,j}
             {{\big( -1 + {\alpha }^2 \big) }^4} + 
           \frac{\big( -2 + j + s \big) \,
          \big( -1 + j + s \big) \,
          \big( -5 + 2\,j + 2\,s \big) 
        \,\big( -3 + 2\,j + 2\,s \big) }
       {{\big( 1 + {\alpha }^2 \big) }^4}
 + \frac{6\,j\,\big( -2 + j + s \big)
 \,\big( -5 + 2\,j + 2\,s \big) }
            {{\big( -1 + {\alpha }^4 \big) }^2} \Big) }
    {2\,\big( -5 + 2\,j + 2\,s \big) } \Big) \Bigg)
\nonumber\\
&\Big{/}&\Big(45\,{\big( 1 + {\alpha }^2 \big) }^4\Big)\,,
\nonumber\\
 w_3 (\alpha,j,s) &=&
\big( -5 + 2\,j + 2\,s \big)
 \,\Big\{ 
16\,\big( -2 + j + s \big) 
\,\big( -1 + j + s \big)
 \,\big( j + s \big) \,
 \big( -3 + 2\,j + 2\,s \big) 
 \,\big( -1 + 2\,j + 2\,s \big)  
\nonumber\\
&+& 
\Big(21\,{\big( 1 + {\alpha }^2 \big) }^6\,
\Big( \frac{15\,\left( -2 + j \right) \,\left( -1 + j \right) \,j}
            {{\left( -1 + {\alpha }^2 \right) }^6} 
\nonumber\\
&+& \frac{\big( -2 + j + s \big) 
            \,\big( -1 + j + s \big) 
             \,\big( j + s \big) \,
              \big( -5 + 2\,j + 2\,s \big) 
          \,\big( -3 + 2\,j + 2\,s \big) 
   \,\big( -1 + 2\,j + 2\,s \big) }
    {{\big( 1 + {\alpha }^2 \big) }^6}
\nonumber\\
&+&  \frac{15\,j\,\big( -2 + j + s \big) 
     \,\big( -1 + j + s \big) 
     \,\big( -5 + 2\,j + 2\,s \big) 
      \,\big( -3 + 2\,j + 2\,s \big) }
 {{\big( -1 + {\alpha }^2 \big) }^2
\,{\big( 1 + {\alpha }^2 \big) }^4}
\nonumber \\
& +& \frac{45\,\big( -1 + j \big) \,j\,
             \big( -2 + j + s \big) \,
             \big( -5 + 2\,j + 2\,s \big)}
        {{\left( -1 + {\alpha }^2 \right) }^4
\,{\big( 1 + {\alpha }^2 \big) }^2} \Big) \Big)/
  \Big(2\,\big( -5 + 2\,j + 2\,s \big) \Big)
\nonumber \\
& -&  \Big(63\,\big( -2 + j + s \big) 
     \,\Big( 4\,\big( -1 + j + s \big) 
     \,\big( j + s \big)
     \,\big( -3 + 2\,j + 2\,s \big) \,
       \big( -1 + 2\,j + 2\,s \big) 
\nonumber\\
& +& \frac{40\,j\,\big( -1 + j + s \big) 
     \,\big( -3 + 2\,j + 2\,s \big) \,  
       {\big( 1 + {\alpha }^2 \big) }^2}
      {{\big( -1 + {\alpha }^2 \big) }^2} 
+ \frac{60\,\big( -1 + j \big) \,j\,
        {\big( 1 + {\alpha }^2 \big) }^4}
        {{\big( -1 + {\alpha }^2 \big) }^4} \Big) \Big)
\Big{/}4 
\nonumber\\
&+& 105\,\big( -2 + j + s \big)
     \,{\big( 1 + {\alpha }^2 \big) }^4\,
       \Big( \frac{3\,\left( -1 + j \right) \,j}
                   {{\left( -1 + {\alpha }^2 \right) }^4}
       + \frac{\big( -1 + j + s \big)
             \,\big( j + s \big) 
             \,\big( -3 + 2\,j + 2\,s \big) 
             \,\big( -1 + 2\,j + 2\,s \big) }
          {{\big( 1 + {\alpha }^2 \big) }^4} 
\nonumber\\
&+& \frac{6\,j\,\big( -1 + j + s \big) 
              \,\big( -3 + 2\,j + 2\,s \big) }
         {{\big( -1 + {\alpha }^4 \big) }^2} \Big)
\nonumber\\  
 &-& \Big(84\,\big( -2 + j + s \big) 
              \,\big( -1 + j + s \big) \,
               \big( -3 + 2\,j + 2\,s \big) 
         \,\Big( 2\,j^2\,{\left( -1 + {\alpha }^2 \right) }^2 
  \nonumber\\
    &+& s\,\big( -1 + 2\,s \big)
      \,{\big( -1 + {\alpha }^2 \big) }^2 
     + 4\,j\,\big( {\alpha }^2 + s\,{\big( -1 + {\alpha }^2 \big) }^2 \big) 
           \Big) \Big)
  \Big{/}
       {{\big( -1 + {\alpha }^2 \big) }^2} \Big\} 
\nonumber\\
&\Big{/}&
  945\,{\big( 1 + {\alpha }^2 \big) }^6\,.
\end{eqnarray}
Thus, we obtain the analytically continued result of
the derivative of the subtracted zeta function given by Eq. (\ref{P})
\begin{eqnarray}
P'(0)
&=& \pi^2 g^4 \sum_{n=1}^{\infty}
 \Bigg( 16 
\Big\{
\Big(\frac{3}{2}-2\ln(2g)\Big)
  \Big[\zeta_{H}(-4,\frac{1}{2}+\frac{1+\alpha^2}{2\kappa}n)
  -F(n,0,0)
\nonumber\\
&+&2 \zeta_{H}{}' (-4,\frac{1}{2}+\frac{1+\alpha^2}{2\kappa}n)
  -\partial_s F(n,0,0)
\Big\}
\nonumber\\
&-& 8\Big\{
   \Big(1-2\ln(2g)\Big)
 \Big[
      \big(\frac{n^2}{\kappa^2}(1-\alpha^2)^2+1\big)
    \zeta_{H} (-2,\frac{1}{2}+\frac{1+\alpha^2}{2\kappa}n)
   - F(n,1,0)  
 \Big]
\nonumber\\
&+&2 \big(\frac{n^2}{\kappa^2}(1-\alpha^2)^2+1\big)
\zeta_{H}{}' (-2,\frac{1}{2}+\frac{1+\alpha^2}{2\kappa}n)
  -\partial_s F(n,1,0)
\Big\}
\nonumber \\
&+&
   \Big(-2\ln(2g)\Big)
 \Big[
      \big(\frac{n^2}{\kappa^2}(1-\alpha^2)^2+1\big)^2
    \zeta_{H} (0,\frac{1}{2}+\frac{1+\alpha^2}{2\kappa}n)
   - F(n,2,0)  
 \Big]
\nonumber\\
&+&2 \big(\frac{n^2}{\kappa^2}(1-\alpha^2)^2+1\big)^2
\zeta_{H}{}' (0,\frac{1}{2}+\frac{1+\alpha^2}{2\kappa}n)
  -\partial_s F(n,2,0)
\nonumber\\
&+&\sum_{j=0}^{\infty}
   \frac{2^{5-2j}}{j(j-1)(j-2)}
   \Big[
    \big(\frac{n^2}{\kappa^2}(1-\alpha^2)^2+1\big)^j
    \zeta_{H} (2j-4,\frac{1}{2}+\frac{1+\alpha^2}{2\kappa}n)
   - F(n,j,0)  
     \Big]
 \Bigg)
\nonumber\\
&=& \pi^2 g^4 \sum_{n=1}^{\infty}
 \Bigg( 16 
\Big\{
\Big(\frac{3}{2}-2\ln(2g)\Big)
 \zeta_{H}(-4,\frac{1}{2}+\frac{1+\alpha^2}{2\kappa}n)
 +2 \zeta_{H}{}' (-4,\frac{1}{2}+\frac{1+\alpha^2}{2\kappa}n)
\nonumber\\
&+&\frac{(1+\alpha^2)^5}{160(\frac{n}{\kappa})}
  \Big(\frac{19}{10}
      +2\ln\Big(\frac{\kappa}{n(1+\alpha^2)g}\Big) 
  \Big)
\nonumber \\
&\times&
\Big(
w_0 (\alpha,0,0) \big(\frac{n}{\kappa}\big)^6
+w_1 (\alpha,0,0) \big(\frac{n}{\kappa}\big)^4
+w_2 (\alpha,0,0) \big(\frac{n}{\kappa}\big)^2
+w_3 (\alpha,0,0) 
\Big)
\nonumber\\
&+&\frac{(1+\alpha^2)^5}{160(\frac{n}{\kappa})}
\Big( \partial_s w_0 (\alpha,0,0) \big(\frac{n}{\kappa}\big)^6
+\partial_s w_1 (\alpha,0,0) \big(\frac{n}{\kappa}\big)^4
+\partial_s w_2 (\alpha,0,0) \big(\frac{n}{\kappa}\big)^2
+\partial_s w_3 (\alpha,0,0) 
\Big)
\Big\}
\nonumber\\
&-& 8\Big\{
\big(\frac{n^2}{\kappa^2}(1-\alpha^2)^2+1\big)
\Big( 
 \big(1-2\ln(2g)\big)
 \zeta_{H}(-2,\frac{1}{2}+\frac{1+\alpha^2}{2\kappa}n)
 +2   \zeta_{H}{}' (-2,\frac{1}{2}+\frac{1+\alpha^2}{2\kappa}n)
\Big)
\nonumber\\
&+&\frac{(1+\alpha^2)^3(1-\alpha^2)^2}{24(\frac{n}{\kappa})}
  \Big(\frac{5}{3}
      +2\ln\Big(\frac{\kappa}{n(1+\alpha^2)g}\Big) 
  \Big)
\nonumber\\
&\times&
\Big( w_0 (\alpha,1,0) \big(\frac{n}{\kappa}\big)^6
+w_1 (\alpha,1,0) \big(\frac{n}{\kappa}\big)^4
+w_2 (\alpha,1,0) \big(\frac{n}{\kappa}\big)^2
+w_3 (\alpha,1,0) 
\Big)
\nonumber\\
&+&\frac{(1+\alpha^2)^3(1-\alpha^2)^2}{24(\frac{n}{\kappa})}
\Big( \partial_s w_0 (\alpha,1,0) \big(\frac{n}{\kappa}\big)^6
+\partial_s w_1 (\alpha,1,0) \big(\frac{n}{\kappa}\big)^4
+\partial_s w_2 (\alpha,1,0) \big(\frac{n}{\kappa}\big)^2
+\partial_s w_3 (\alpha,1,0) 
\Big)
\Big\}
\nonumber \\
&+&
\big(\frac{n^2}{\kappa^2}(1-\alpha^2)^2+1\big)
\Big( 
 -2\ln(2g)
 \zeta_{H}(0,\frac{1}{2}+\frac{1+\alpha^2}{2\kappa}n)
 +2   \zeta_{H}{}' (0,\frac{1}{2}+\frac{1+\alpha^2}{2\kappa}n)
\Big)
\nonumber\\
&+&\frac{(1+\alpha^2)(1-\alpha^2)^4}{2(\frac{n}{\kappa})}
  \Big(2
      +2\ln\Big(\frac{\kappa}{n(1+\alpha^2)g}\Big) 
  \Big)
\nonumber\\
&\times&
\Big( w_0 (\alpha,2,0) \big(\frac{n}{\kappa}\big)^6
+w_1 (\alpha,2,0) \big(\frac{n}{\kappa}\big)^4
+w_2 (\alpha,2,0) \big(\frac{n}{\kappa}\big)^2
+w_3 (\alpha,2,0) 
\Big)
\nonumber\\
&+&\frac{(1+\alpha^2)(1-\alpha^2)^4}{2(\frac{n}{\kappa})}
\Big( \partial_s w_0 (\alpha,2,0) \big(\frac{n}{\kappa}\big)^6
+\partial_s w_1 (\alpha,2,0) \big(\frac{n}{\kappa}\big)^4
+\partial_s w_2 (\alpha,2,0) \big(\frac{n}{\kappa}\big)^2
+\partial_s w_3 (\alpha,2,0) 
\Big)
\nonumber\\
&+&\sum_{j=3}^{\infty}
   \frac{2^{5-2j}}{j(j-1)(j-2)}
   \Big[
    \big(\frac{n^2}{\kappa^2}(1-\alpha^2)^2+1\big)^j
    \zeta_{H} (2j-4,\frac{1}{2}+\frac{1+\alpha^2}{2\kappa}n)
\nonumber\\
&-& \frac{2^{-5+2j}(1-\alpha^2)^{2j}}
      {(2j-5)(1+\alpha^2)^{2j-5}(\frac{n}{\kappa})}
\Big( w_0 (\alpha,j,0) \big(\frac{n}{\kappa}\big)^6
+ w_1 (\alpha,j,0) \big(\frac{n}{\kappa}\big)^4
+w_2 (\alpha,j,0) \big(\frac{n}{\kappa}\big)^2
+ w_3 (\alpha,j,0) 
\Big)
     \Big]
 \Bigg)
\,.
\label{PD6D}
\end{eqnarray}

Next we focus on the quantities appearing in the
counter terms.
The analytic continuation of the $j=0$ term is given by 
\begin{eqnarray}
\frac{d}{ds} 
\Big(G(0,s)\Delta(0,s)\Big)_{s\to 0}
&=&
-\frac{\kappa(1+\alpha^2)^5\pi^2 g^4}
      {10}
\nonumber\\
&\times& \Big[
\Big(
 2\ln\Big(\frac{\kappa}{g^2(1+\alpha^2)}\Big)
+\frac{19}{10}
\Big)
\Big(
-\frac{w_0(\alpha,0,0)}{252\kappa^6}
+\frac{w_1(\alpha,0,0)}{120\kappa^4}
-\frac{w_2(\alpha,0,0)}{12\kappa^2}
\nonumber\\
&+&\big(w_3(\alpha,0,s)\zeta(2s+1)\big)_{s\to 0}
\Big)
\nonumber\\
&+&\frac{1}{120}\frac{w_1{}'(\alpha,0,0))}
                     {\kappa^4}
 -\frac{1}{12}\frac{w_2{}'(\alpha,0,0))}
                     {\kappa^2}
+\frac{2w_0(\alpha,0,0)}{\kappa^6}
    \zeta_R'(-5)
+\frac{2w_1(\alpha,0,0)}{\kappa^4}
    \zeta_R'(-3)
\nonumber\\
&+&\frac{2w_2(\alpha,0,0)}{\kappa^2}
    \zeta_R'(-1)
+\partial_s\big(w_3(\alpha,0,s)\zeta_R(2s+1)\big)_{s\to 0}
\Big]\,,
\end{eqnarray}
where we have used
\begin{eqnarray}
w_3(\alpha,0,s)\zeta_R(2s+1)
=\frac{31}{126\,{\left( 1 + {\alpha }^2 \right) }^6} - \frac{31\,\left( \frac{137}{2} - 30\,\gamma \right) \,s}
   {1890\,{\left( 1 + {\alpha }^2 \right) }^6} 
+O(s^2)\,.
\end{eqnarray}
Similarly, contributions from $j=1$ and $j=2$ are given by
respectively
\begin{eqnarray}
\frac{d}{ds} 
\Big(G(1,s)\Delta(1,s)\Big)_{s\to 0}
&=&
\frac{\kappa(1+\alpha^2)^3(1-\alpha^2)^2\pi^2 g^4}
      {3}
\nonumber\\
&\times& \Big[
\Big(
 2\ln\Big(\frac{\kappa}{g^2(1+\alpha^2)}\Big)
+\frac{5}{3}
\Big)
\Big(
-\frac{w_0(\alpha,1,0)}{252\kappa^6}
+\frac{w_1(\alpha,1,0)}{120\kappa^4}
-\frac{w_2(\alpha,1,0)}{12\kappa^2}
\nonumber\\
&+&\big(w_3(\alpha,1,s)\zeta(2s+1)\big)_{s\to 0}
\Big)
\nonumber\\
&+&\frac{1}{120}\frac{w_1{}'(\alpha,1,0))}
                     {\kappa^4}
 -\frac{1}{12}\frac{w_2{}'(\alpha,1,0))}
                     {\kappa^2}
+\frac{2w_0(\alpha,1,0)}{\kappa^6}
    \zeta_R'(-5)
+\frac{2w_1(\alpha,1,0)}{\kappa^4}
    \zeta_R'(-3)
\nonumber\\
&+&\frac{2w_2(\alpha,1,0)}{\kappa^2}
    \zeta_R'(-1)
+\partial_s\big(w_3(\alpha,1,s)\zeta_R(2s+1)\big)_{s\to 0}
\Big]\,,
\end{eqnarray}
where
\begin{eqnarray}
w_3(\alpha,1,s)\zeta_R(2s+1)
&=&\frac{-\left( 29 + 89\,{\alpha }^2 + 29\,{\alpha }^4 \right) }
{315\,{\left(-1+{\alpha }^2\right) }^2\,
{\left( 1 + {\alpha }^2 \right) }^6} 
\nonumber\\
&+& \frac{\left( 1555 + 3358\,{\alpha }^2 + 1555\,{\alpha }^4 
- 24\,\gamma \,\left( 29 + 89\,{\alpha }^2
 + 29\,{\alpha }^4 \right)  \right) \,s}
{3780\,{\left( -1 + {\alpha }^2 \right) }^2\,
{\left( 1 + {\alpha }^2 \right) }^6} + O(s^2)\,,
\end{eqnarray}
and
\begin{eqnarray}
\frac{d}{ds} 
\Big(G(2,s)\Delta(2,s)\Big)_{s\to 0}
&=&
-\frac{\kappa(1+\alpha^2) (1-\alpha^2)^4\pi^2 g^4}
      {2}
\nonumber\\
&\times& \Big[
\Big(
 2\ln\Big(\frac{\kappa}{g^2(1+\alpha^2)}\Big)
+2
\Big)
\Big(
-\frac{w_0(\alpha,2,0)}{252\kappa^6}
+\frac{w_1(\alpha,2,0)}{120\kappa^4}
-\frac{w_2(\alpha,2,0)}{12\kappa^2}
\nonumber\\
&+&\big(w_3(\alpha,2,0)\zeta(2s+1)\big)_{s\to 0}
\Big)
\nonumber\\
&+&\frac{1}{120}\frac{w_1{}'(\alpha,2,0))}
                     {\kappa^4}
 -\frac{1}{12}\frac{w_2{}'(\alpha,2,0))}
                     {\kappa^2}
+\frac{2w_0(\alpha,2,0)}{\kappa^6}
    \zeta_R'(-5)
+\frac{2w_1(\alpha,2,0)}{\kappa^4}
    \zeta_R'(-3)
\nonumber\\
&+&\frac{2w_2(\alpha,2,0)}{\kappa^2}
    \zeta_R'(-1)
+\partial_s\big(w_3(\alpha,2,s)\zeta_R(2s+1)\big)_{s\to 0}
\Big]\,,
\end{eqnarray}
where
\begin{eqnarray}
&&w_3(\alpha,2,s)\zeta_R(2s+1)
\nonumber\\
&=&\frac{87 + 296\,{\alpha }^2 + 914\,{\alpha }^4
       + 296\,{\alpha }^6+ 87\,{\alpha }^8}
   {630\,{\left( -1 + {\alpha }^2 \right) }^4\,
     {\left( 1 + {\alpha }^2 \right)}^6} 
\nonumber\\
&+&  \frac{\left( -1151 + 1044\,\gamma 
+ 4\,\left( -1663 + 888\,\gamma \right) \,{\alpha }^2 + 
       6\,\left( -759 + 1828\,\gamma \right) \,{\alpha }^4 + 4\,\left( -1663 + 888\,\gamma \right) \,{\alpha }^6 + 
       \left( -1151 + 1044\,\gamma \right) \,{\alpha }^8 \right) \,s}
{3780\,{\left( -1 + {\alpha }^2 \right) }^4\,
     {\left( 1 + {\alpha }^2 \right) }^6}
\nonumber\\
& +& O(s^2)\,.
\end{eqnarray}

Finally, for the contribution from $j\geq 3$, we obtain
\begin{eqnarray}
\sum_{j=3}^{\infty}
\frac{d}{ds}\Big(G(j,s)\Delta(j,s)\Big)_{s\to 0}
&=&
\sum_{j=3}^{\infty}\frac{\kappa(1+\alpha^2)^{5-2j}
       (1-\alpha^2)^{2j}\pi^2 g^4}
      {(2j-5)j(j-1)(j-2)}
\nonumber\\
&\times& 
\Big[
-\frac{w_0(\alpha,j,0)}{252\kappa^6}
+\frac{w_1(\alpha,j,0)}{120\kappa^4}
-\frac{w_2(\alpha,j,0)}{12\kappa^2}
\nonumber\\
&+&\Big(
 -\frac{2}{2j-5}
 +2\ln\Big(\frac{\kappa}{g^2(1+\alpha^2)}
 +\psi(j-2)
   \Big)
\Big(\frac{w_3(\alpha,j,s)\zeta_R(2s+1)}
          {\Gamma(s)}
\Big)_{s\to 0}
\nonumber\\
&+&
\partial_s
\Big(\frac{w_3(\alpha,j,s)\zeta_R(2s+1)}
          {\Gamma(s)}
\Big)_{s\to 0}
\Big]\,.
\end{eqnarray}

Applying similar scheme of analytic continuation, we obtain
\begin{eqnarray}
&&\sum_{j=0}^{\infty}G(j,0)\Delta(j,0)=
\nonumber\\
&-&\frac{\kappa(1+\alpha^2)^5\pi^2 g^4}
      {10}
\Big(
-\frac{w_0(\alpha,0,0)}{252\kappa^6}
+\frac{w_1(\alpha,0,0)}{120\kappa^4}
-\frac{w_2(\alpha,0,0)}{12\kappa^2}
+\big(w_3(\alpha,0,s)\zeta(2s+1)\big)_{s\to 0}
\Big)
\nonumber \\
&+&\frac{\kappa(1+\alpha^2)^3 (1-\alpha^2)^2\pi^2 g^4}
      {3}
\Big(
-\frac{w_0(\alpha,1,0)}{252\kappa^6}
+\frac{w_1(\alpha,1,0)}{120\kappa^4}
-\frac{w_2(\alpha,1,0)}{12\kappa^2}
+\big(w_3(\alpha,1,s)\zeta(2s+1)\big)_{s\to 0}
\Big)
\nonumber\\
&-&\frac{\kappa(1+\alpha^2) (1-\alpha^2)^4\pi^2 g^4}
      {2}
\Big(
-\frac{w_0(\alpha,2,0)}{252\kappa^6}
+\frac{w_1(\alpha,2,0)}{120\kappa^4}
-\frac{w_2(\alpha,2,0)}{12\kappa^2}
+\big(w_3(\alpha,2,s)\zeta(2s+1)\big)_{s\to 0}
\Big)
\nonumber\\
&+&
\sum_{j=3}^{\infty}\frac{\kappa(1+\alpha^2)^{5-2j}
       (1-\alpha^2)^{2j}\pi^2 g^4}
      {(2j-5)j(j-1)(j-2)}
\Big(\frac{w_3(\alpha,j,s)\zeta_R(2s+1)}
          {\Gamma(s)}
\Big)_{s\to 0}\,.
\end{eqnarray}


\section{Zeta functions for axisymmetric Kaluza-Klein modes}

We now derive the derivative of zeta functions for axisymmetric modes.
The axisymmetric zeta function can be written as
\begin{eqnarray}
(2\pi)^4\zeta_0(s)
=\int d^4x \sum_{j=0}^{\infty}
 \frac{2^{5-2s-2j} \pi^2 g^{2(2-s)}\Gamma(j+s-2)} 
      {j!\Gamma(s)}
 \zeta_H(2s+2j-4,\frac{1}{2})\,.
\end{eqnarray}
After taking derivative with respect to $s$,
 analytic continuation to $s\to 0$ gives
\begin{eqnarray}
(2\pi)^4\zeta_0{}'(0)
=\int d^4 x\,\, g^4
\Big(-8\zeta_R(3)
   +24\frac{\zeta_R(5)}{\pi^2}
\Big)\,,
\end{eqnarray}
where we have used
\begin{eqnarray}
\sum_{j=3}^{\infty}
\frac{2^{5-2j}}{j(j-1)(j-2)}
  \zeta_H(2j-4,\frac{1}{2})
=\ln2 -\frac{5\zeta_R(3)}{\pi^2}
  +\frac{93\zeta_R(5)}{2\pi^4}\,.
\end{eqnarray}
and
$\zeta_{R}(z)= 
 \pi^{z-1/2}\zeta_R(1-z)\Gamma((1-z)/2)/\Gamma(z/2)$.
The analytic continuation of zeta function itself is
\begin{eqnarray}
(2\pi)^4\zeta_0(0)=0\,.
\end{eqnarray}

\section{Functions related to the  heat kernel coefficient}

The function ${\cal F}(z)$ in Eq. (\ref{aa6}) is defined by
\begin{eqnarray}
{\cal F}(z)&:=&
  1748+4964\,z^2+5461\,z^4+2035\,z^6 - 1740\,{\alpha }^2 
- 5936\,z^2\,{\alpha }^2 - 9049\,z^4\,{\alpha }^2 - 2882\,z^6\,{\alpha }^2 
\nonumber\\
&+& 711\,z^8\,{\alpha }^2 + 1564\,{\alpha }^4 + 13944\,z^2\,{\alpha }^4 + 32534\,z^4\,{\alpha }^4 + 26109\,z^6\,{\alpha }^4 + 12757\,z^8\,{\alpha }^4 
\nonumber\\
&+& 3652\,z^{10}\,{\alpha }^4 - 548\,{\alpha }^6 - 10480\,z^2\,{\alpha }^6
 - 27054\,z^4\,{\alpha }^6 -  30044\,z^6\,{\alpha }^6 - 27054\,z^8\,{\alpha }^6
\nonumber\\
&-& 10480\,z^{10}\,{\alpha }^6 - 548\,z^{12}\,{\alpha }^6 
  + 3652\,z^2\,{\alpha }^8 +  12757\,z^4\,{\alpha }^8 + 26109\,z^6\,{\alpha }^8 + 32534\,z^8\,{\alpha }^8 + 13944\,z^{10}\,{\alpha }^8 
\nonumber\\
&+ &1564\,z^{12}\,{\alpha }^8+711\,z^4\,{\alpha }^{10}
- 2882\,z^6\,{\alpha }^{10} - 9049\,z^8\,{\alpha }^{10}
 - 5936\,z^{10}\,{\alpha }^{10} - 1740\,z^{12}\,{\alpha }^{10}  
\nonumber\\
&+& 2035\,z^6\,{\alpha }^{12} + 5461\,z^8\,{\alpha }^{12} + 4964\,z^{10}\,{\alpha }^{12} + 1748\,z^{12}\,{\alpha }^{12}\,.
\end{eqnarray}
The function ${\cal G}(\epsilon, z)$ in Eq.~(\ref{bb6}) is also defined by
\begin{eqnarray}
{\cal G}(\epsilon,z)
&:=&
1024 + 2580\,z^2 + 2808\,z^4 + 1042\,z^6 + 4176\,z^2\,{\alpha }^2 
+ 10176\,z^4\,{\alpha }^2 + 11964\,z^6\,{\alpha }^2 + 4704\,z^8\,{\alpha }^2
\nonumber\\
 &-&4392\,z^2\,{\alpha }^4 - 15984\,z^4\,{\alpha }^4 - 26130\,z^6\,{\alpha }^4
- 11976\,z^8\,{\alpha }^4 - 588\,z^{10}\,{\alpha }^4 + 4368\,z^2\,{\alpha }^6 
\nonumber\\
&+&  25632\,z^4\,{\alpha }^6 + 46728\,z^6\,{\alpha }^6 + 25632\,z^8\,{\alpha }^6+ 4368\,z^{10}\,{\alpha }^6 - 588\,z^2\,{\alpha }^8 - 11976\,z^4\,{\alpha }^8 
\nonumber\\
&-& 26130\,z^6\,{\alpha }^8 - 15984\,z^8\,{\alpha }^8
 - 4392\,z^{10}\,{\alpha }^8 + 4704\,z^4\,{\alpha }^{10}
 + 11964\,z^6\,{\alpha }^{10} + 10176\,z^8\,{\alpha }^{10}
 +  4176\,z^{10}\,{\alpha }^{10} 
\nonumber\\
&+&
 1042\,z^6\,{\alpha }^{12} + 2808\,z^8\,{\alpha }^{12} + 2580\,z^{10}\,{\alpha }^{12} 
+ 1024\,z^{12}\,{\alpha }^{12} + 704\,\epsilon  +   2368\,z^2\,\epsilon 
 + 2678\,z^4\,\epsilon  + 1014\,z^6\,\epsilon
\nonumber\\
& -& 1728\,{\alpha }^2\,\epsilon  - 10272\,z^2\,{\alpha }^2\,\epsilon  - 
  19726\,z^4\,{\alpha }^2\,\epsilon  - 15332\,z^6\,{\alpha }^2\,\epsilon  - 4150\,z^8\,{\alpha }^2\,\epsilon  + 1600\,{\alpha }^4\,\epsilon  
\nonumber\\
&+&  18912\,z^2\,{\alpha }^4\,\epsilon  + 50460\,z^4\,{\alpha }^4\,\epsilon  + 54762\,z^6\,{\alpha }^4\,\epsilon  + 26062\,z^8\,{\alpha }^4\,\epsilon  + 
  4448\,z^{10}\,{\alpha }^4\,\epsilon  - 576\,{\alpha }^6\,\epsilon 
\nonumber\\
&-& 15456\,z^2\,{\alpha }^6\,\epsilon  - 55324\,z^4\,{\alpha }^6\,\epsilon 
 -  80888\,z^6\,{\alpha }^6\,\epsilon  - 55324\,z^8\,{\alpha }^6\,\epsilon  
- 15456\,z^{10}\,{\alpha }^6\,\epsilon  - 576\,z^{12}\,{\alpha }^6\,\epsilon  
\nonumber\\
&+& 4448\,z^2\,{\alpha }^8\,\epsilon  + 26062\,z^4\,{\alpha }^8\,\epsilon  + 54762\,z^6\,{\alpha }^8\,\epsilon  + 50460\,z^8\,{\alpha }^8\,\epsilon  + 
  18912\,z^{10}\,{\alpha }^8\,\epsilon  + 1600\,z^{12}\,{\alpha }^8\,\epsilon 
\nonumber\\
&-& 4150\,z^4\,{\alpha }^{10}\,\epsilon  - 15332\,z^6\,{\alpha }^{10}\,\epsilon -19726\,z^8\,{\alpha }^{10}\,\epsilon  - 10272\,z^{10}\,{\alpha }^{10}\,\epsilon -1728\,z^{12}\,{\alpha }^{10}\,\epsilon  + 1014\,z^6\,{\alpha }^{12}\,\epsilon \nonumber\\
& +& 
  2678\,z^8\,{\alpha }^{12}\,\epsilon  + 2368\,z^{10}\,{\alpha }^{12}\,\epsilon  + 704\,z^{12}\,{\alpha }^{12}\,\epsilon  + 32\,{\epsilon }^2 + 
  152\,z^2\,{\epsilon }^2 + 204\,z^4\,{\epsilon }^2 + 84\,z^6\,{\epsilon }^2 
\nonumber\\
&- &48\,{\alpha }^2\,{\epsilon }^2 - 464\,z^2\,{\alpha }^2\,{\epsilon }^2 - 
  1028\,z^4\,{\alpha }^2\,{\epsilon }^2 - 872\,z^6\,{\alpha }^2\,{\epsilon }^2 - 260\,z^8\,{\alpha }^2\,{\epsilon }^2 + 480\,z^2\,{\alpha }^4\,{\epsilon }^2 
\nonumber\\
&+& 
  1688\,z^4\,{\alpha }^4\,{\epsilon }^2 + 1964\,z^6\,{\alpha }^4\,{\epsilon }^2 + 764\,z^8\,{\alpha }^4\,{\epsilon }^2 + 8\,z^{10}\,{\alpha }^4\,{\epsilon }^2 + 
  16\,{\alpha }^6\,{\epsilon }^2 - 176\,z^2\,{\alpha }^6\,{\epsilon }^2
 - 1368\,z^4\,{\alpha }^6\,{\epsilon }^2 - 2352\,z^6\,{\alpha }^6\,{\epsilon }^2\nonumber\\
&-& 1368\,z^8\,{\alpha }^6\,{\epsilon }^2 - 176\,z^{10}\,{\alpha }^6\,{\epsilon }^2 + 16\,z^{12}\,{\alpha }^6\,{\epsilon }^2 
+ 8\,z^2\,{\alpha }^8\,{\epsilon }^2 +   764\,z^4\,{\alpha }^8\,{\epsilon }^2
 + 1964\,z^6\,{\alpha }^8\,{\epsilon }^2 + 1688\,z^8\,{\alpha }^8\,{\epsilon }^2\nonumber\\
&+& 480\,z^{10}\,{\alpha }^8\,{\epsilon }^2 
-   260\,z^4\,{\alpha }^{10}\,{\epsilon }^2 
- 872\,z^6\,{\alpha }^{10}\,{\epsilon }^2
 - 1028\,z^8\,{\alpha }^{10}\,{\epsilon }^2 
- 464\,z^{10}\,{\alpha }^{10}\,{\epsilon }^2
 - 48\,z^{12}\,{\alpha }^{10}\,{\epsilon }^2
\nonumber\\
& + &84\,z^6\,{\alpha }^{12}\,{\epsilon }^2 
+  204\,z^8\,{\alpha }^{12}\,{\epsilon }^2 
+  152\,z^{10}\,{\alpha }^{12}\,{\epsilon }^2 
+ 32\,z^{12}\,{\alpha }^{12}\,{\epsilon }^2
 - 12\,{\epsilon }^3 -   132\,z^2\,{\epsilon }^3 - 222\,z^4\,{\epsilon }^3
 - 102\,z^6\,{\epsilon }^3 + 36\,{\alpha }^2\,{\epsilon }^3
\nonumber\\
& +& 608\,z^2\,{\alpha }^2\,{\epsilon }^3 
+  1490\,z^4\,{\alpha }^2\,{\epsilon }^3
 + 1324\,z^6\,{\alpha }^2\,{\epsilon }^3 + 406\,z^8\,{\alpha }^2\,{\epsilon }^3 - 36\,{\alpha }^4\,{\epsilon }^3 -  1032\,z^2\,{\alpha }^4\,{\epsilon }^3 
- 3544\,z^4\,{\alpha }^4\,{\epsilon }^3 
\nonumber\\
&- &4378\,z^6\,{\alpha }^4\,{\epsilon }^3 - 2042\,z^8\,{\alpha }^4\,{\epsilon }^3 -   212\,z^{10}\,{\alpha }^4\,{\epsilon }^3 + 12\,{\alpha }^6\,{\epsilon }^3 + 768\,z^2\,{\alpha }^6\,{\epsilon }^3 + 3912\,z^4\,{\alpha }^6\,{\epsilon }^3
\nonumber\\
& +& 6312\,z^6\,{\alpha }^6\,{\epsilon }^3 
+ 3912\,z^8\,{\alpha }^6\,{\epsilon}^3 
+ 768\,z^{10}\,{\alpha }^6\,{\epsilon }^3 
+ 12\,z^{12}\,{\alpha }^6\,{\epsilon }^3 
-   212\,z^2\,{\alpha }^8\,{\epsilon }^3
 - 2042\,z^4\,{\alpha }^8\,{\epsilon }^3 
- 4378\,z^6\,{\alpha }^8\,{\epsilon }^3
\nonumber\\
&-& 3544\,z^8\,{\alpha }^8\,{\epsilon }^3 
-   1032\,z^{10}\,{\alpha }^8\,{\epsilon }^3 
- 36\,z^{12}\,{\alpha }^8\,{\epsilon }^3
 + 406\,z^4\,{\alpha }^{10}\,{\epsilon }^3 
+  1324\,z^6\,{\alpha }^{10}\,{\epsilon }^3 
+ 1490\,z^8\,{\alpha }^{10}\,{\epsilon }^3 
+ 608\,z^{10}\,{\alpha }^{10}\,{\epsilon }^3 
\nonumber\\
&+ & 36\,z^{12}\,{\alpha }^{10}\,{\epsilon }^3
- 102\,z^6\,{\alpha }^{12}\,{\epsilon }^3 
- 222\,z^8\,{\alpha }^{12}\,{\epsilon }^3 
-  132\,z^{10}\,{\alpha }^{12}\,{\epsilon }^3 
- 12\,z^{12}\,{\alpha }^{12}\,{\epsilon }^3 
- 4\,z^2\,{\epsilon }^4 - 10\,z^4\,{\epsilon }^4
 - 6\,z^6\,{\epsilon }^4 
\nonumber\\
&+&  16\,z^2\,{\alpha }^2\,{\epsilon }^4 
54\,z^4\,{\alpha }^2\,{\epsilon }^4
+ 52\,z^6\,{\alpha }^2\,{\epsilon }^4 + 14\,z^8\,{\alpha }^2\,{\epsilon }^4 - 
  24\,z^2\,{\alpha }^4\,{\epsilon }^4 - 116\,z^4\,{\alpha }^4\,{\epsilon }^4 - 154\,z^6\,{\alpha }^4\,{\epsilon }^4
\nonumber\\
& -& 66\,z^8\,{\alpha }^4\,{\epsilon }^4 - 
  4\,z^{10}\,{\alpha }^4\,{\epsilon }^4 
+16\,z^2\,{\alpha }^6\,{\epsilon }^4 
+ 124\,z^4\,{\alpha }^6\,{\epsilon }^4 
+ 216\,z^6\,{\alpha }^6\,{\epsilon }^4 
\nonumber\\
&+& 124\,z^8\,{\alpha }^6\,{\epsilon }^4
 + 16\,z^{10}\,{\alpha }^6\,{\epsilon }^4 
- 4\,z^2\,{\alpha }^8\,{\epsilon }^4
 - 66\,z^4\,{\alpha }^8\,{\epsilon }^4 
-  154\,z^6\,{\alpha }^8\,{\epsilon }^4
- 116\,z^8\,{\alpha }^8\,{\epsilon }^4 -
 24\,z^{10}\,{\alpha }^8\,{\epsilon }^4 
+ 14\,z^4\,{\alpha }^{10}\,{\epsilon }^4 
\nonumber\\
&+ &52\,z^6\,{\alpha }^{10}\,{\epsilon }^4
 + 54\,z^8\,{\alpha }^{10}\,{\epsilon }^4 
+ 16\,z^{10}\,{\alpha }^{10}\,{\epsilon }^4 
- 6\,z^6\,{\alpha }^{12}\,{\epsilon }^4 
- 10\,z^8\,{\alpha }^{12}\,{\epsilon }^4 
- 4\,z^{10}\,{\alpha }^{12}\,{\epsilon }^4 + 3\,z^4\,{\epsilon }^5
\nonumber \\
&+& 3\,z^6\,{\epsilon }^5 - 15\,z^4\,{\alpha }^2\,{\epsilon }^5 
- 18\,z^6\,{\alpha }^2\,{\epsilon }^5 - 3\,z^8\,{\alpha }^2\,{\epsilon }^5
 + 30\,z^4\,{\alpha }^4\,{\epsilon }^5 
+ 45\,z^6\,{\alpha }^4\,{\epsilon }^5 + 15\,z^8\,{\alpha }^4\,{\epsilon }^5
 - 30\,z^4\,{\alpha }^6\,{\epsilon }^5
\nonumber\\
&-& 60\,z^6\,{\alpha }^6\,{\epsilon }^5 
-  30\,z^8\,{\alpha }^6\,{\epsilon }^5 + 15\,z^4\,{\alpha }^8\,{\epsilon }^5 + 45\,z^6\,{\alpha }^8\,{\epsilon }^5 + 30\,z^8\,{\alpha }^8\,{\epsilon }^5 - 
  3\,z^4\,{\alpha }^{10}\,{\epsilon }^5 
\nonumber \\
&-& 18\,z^6\,{\alpha }^{10}\,{\epsilon }^5 
- 15\,z^8\,{\alpha }^{10}\,{\epsilon }^5 
+ 3\,z^6\,{\alpha }^{12}\,{\epsilon }^5
 +   3\,z^8\,{\alpha }^{12}\,{\epsilon }^5\,.
\end{eqnarray}

\newpage


\begin{thebibliography}{99}


\bibitem{Gibbons:2003di}
  G.~W.~Gibbons, R.~Guven and C.~N.~Pope,
  Phys.\ Lett.\ B {\bf 595}, 498 (2004)
  [arXiv:hep-th/0307238];
  Y.~Aghababaie {\it et al.},
  JHEP {\bf 0309}, 037 (2003)
  [arXiv:hep-th/0308064];
%
  C.~P.~Burgess, F.~Quevedo, G.~Tasinato and I.~Zavala,
  JHEP {\bf 0411}, 069 (2004)
  [arXiv:hep-th/0408109].





\bibitem{Arkani-Hamed:1998rs}
  N.~Arkani-Hamed, S.~Dimopoulos and G.~R.~Dvali,
  Phys.\ Lett.\  B {\bf 429}, 263 (1998)
  [arXiv:hep-ph/9803315];
  I.~Antoniadis, N.~Arkani-Hamed, S.~Dimopoulos and G.~R.~Dvali,
  Phys.\ Lett.\  B {\bf 436}, 257 (1998)
  [arXiv:hep-ph/9804398].




\bibitem{Kokorelis}
 D.~Cremades, L.~E.~Ibanez and F.~Marchesano,
  Nucl.\ Phys.\  B {\bf 643}, 93 (2002)
  [arXiv:hep-th/0205074];
  C.~Kokorelis,
  Nucl.\ Phys.\  B {\bf 677}, 115 (2004)
  [arXiv:hep-th/0207234].



\bibitem{Carroll:2003db}
  S.~M.~Carroll and M.~M.~Guica,
  arXiv:hep-th/0302067;
  I.~Navarro,
  JCAP {\bf 0309}, 004 (2003)
  [arXiv:hep-th/0302129];
  Y.~Aghababaie, C.~P.~Burgess, S.~L.~Parameswaran and F.~Quevedo,
  Nucl.\ Phys.\  B {\bf 680}, 389 (2004)
  [arXiv:hep-th/0304256];



\bibitem{Randall:1999ee}
  L.~Randall and R.~Sundrum,
  Phys.\ Rev.\ Lett.\  {\bf 83}, 3370 (1999)
  [arXiv:hep-ph/9905221];
  L.~Randall and R.~Sundrum,
  Phys.\ Rev.\ Lett.\  {\bf 83}, 4690 (1999)
  [arXiv:hep-th/9906064].













\bibitem{Garriga:2000jb}
  J.~Garriga, O.~Pujolas and T.~Tanaka,
  Nucl.\ Phys.\  B {\bf 605}, 192 (2001)
  [arXiv:hep-th/0004109];
  W.~D.~Goldberger and I.~Z.~Rothstein,
  Phys.\ Lett.\  B {\bf 491}, 339 (2000)
  [arXiv:hep-th/0007065];
A.~Flachi, I.~G.~Moss and D.~J.~Toms,
  Phys.\ Rev.\  D {\bf 64}, 105029 (2001)
  [arXiv:hep-th/0106076];
 J.~Garriga, O.~Pujolas and T.~Tanaka,
  Nucl.\ Phys.\  B {\bf 655}, 127 (2003)
  [arXiv:hep-th/0111277];
W.~Naylor and M.~Sasaki,
  Phys.\ Lett.\  B {\bf 542}, 289 (2002)
  [arXiv:hep-th/0205277];
  A.~A.~Saharian and M.~R.~Setare,
  Phys.\ Lett.\  B {\bf 552}, 119 (2003)
  [arXiv:hep-th/0207138];
  I.~G.~Moss and J.~P.~Norman,
  JHEP {\bf 0409}, 020 (2004)
  [arXiv:hep-th/0401181].



\bibitem{Chen:2006nu}
  P.~Chen,
  arXiv:hep-ph/0611378.


\bibitem{cc2}
I.~Navarro,
  Class.\ Quant.\ Grav.\  {\bf 20}, 3603 (2003)
  [arXiv:hep-th/0305014];
  H.~P.~Nilles, A.~Papazoglou and G.~Tasinato,
  Nucl.\ Phys.\  B {\bf 677}, 405 (2004)
  [arXiv:hep-th/0309042];
  H.~M.~Lee,
  Phys.\ Lett.\  B {\bf 587}, 117 (2004)
  [arXiv:hep-th/0309050];
  J.~Vinet and J.~M.~Cline,
  Phys.\ Rev.\  D {\bf 70}, 083514 (2004)
  [arXiv:hep-th/0406141];
J.~Garriga and M.~Porrati,
  JHEP {\bf 0408}, 028 (2004)
  [arXiv:hep-th/0406158].










\bibitem{Nishino:1984gk}
  H.~Nishino and E.~Sezgin,
  Phys.\ Lett.\  B {\bf 144}, 187 (1984).

\bibitem{Salam:1984cj}
  A.~Salam and E.~Sezgin,
  Phys.\ Lett.\  B {\bf 147}, 47 (1984).







\bibitem{mns}
  M.~Minamitsuji, W.~Naylor and M.~Sasaki,
  JHEP {\bf 0612}, 079 (2006)
  [arXiv:hep-th/0606238].


\bibitem{emn}
  E.~Elizalde, M.~Minamitsuji and W.~Naylor,
  Phys.\ Rev.\  D {\bf 75}, 064032 (2007)
  [arXiv:hep-th/0702098].

\bibitem{VSV}
  D.~V.~Vassilevich,
  Phys.\ Rept.\  {\bf 388}, 279 (2003)
  [arXiv:hep-th/0306138].



\bibitem{Milton}
  K.~A.~Milton, S.~D.~Odintsov and S.~Zerbini,
  Phys.\ Rev.\  D {\bf 65}, 065012 (2002)
  [arXiv:hep-th/0110051].






\bibitem{Carter:2006uk}
  B.~M.~N.~Carter, A.~B.~Nielsen and D.~L.~Wiltshire,
  JHEP {\bf 0607}, 034 (2006)
  [arXiv:hep-th/0602086].

\bibitem{Parameswaran:2006db}
  S.~L.~Parameswaran, S.~Randjbar-Daemi and A.~Salvio,
  Nucl.\ Phys.\  B {\bf 767}, 54 (2007)
  [arXiv:hep-th/0608074].











\bibitem{Hoover}
  C.~P.~Burgess and D.~Hoover,
  arXiv:hep-th/0504004;
  D.~Hoover and C.~P.~Burgess,
  JHEP {\bf 0601}, 058 (2006)
  [arXiv:hep-th/0507293];





 \bibitem{ElizaldeCS}
  E.~Elizalde,
  Commun.\ Math.\ Phys.\  {\bf 198} (1998) 83
  [arXiv:hep-th/9707257]; J. Comput.
Appl. Math. {\bf 118}, 125 (2000); J. Phys. {\bf A34}, 3025  (2001);
E. Elizalde, S. Nojiri, S.~D. Odintsov and S. Ogushi, Phys. Rev.
{\bf D67}, 063515 (2003).





\bibitem{Frank:2007jb}
  M.~Frank, I.~Turan and L.~Ziegler,
  arXiv:0704.3626 [hep-ph].






\bibitem{ElizaldeBook}
  E.~Elizalde,
  {\it Ten physical applications of spectral zeta functions},
  Lect.\ Notes Phys.\  {\bf M35} (Springer, Berlin, 1995);
E.~Elizalde et al., {\it Zeta regularization techniques with applications}
(World Scientific, Singapore, 1994)




\end{thebibliography}
\end{document}